\documentclass[12pt]{article}

\def\ds{\displaystyle}
\newcommand{\sss}[1]{\scriptscriptstyle{#1}}

\def\bq{\begin{equation}}
\def\eq{\end{equation}}
\def\bqa{\begin{eqnarray}}
\def\eqa{\end{eqnarray}}
\def\nll{\nonumber\\}
\def\ieps{i \varepsilon}
\def\mz {M_{\sss{Z}}}
\def\mw {M_{\sss{W}}}

\def\mup{m_u}
\def\mdn{m_d}
\def\mgm{m_\gamma}

\def\Pl{P_1^2}
\def\Pll{P_2^2}
\def\Ps{P^2}
\def\Qs{Q^2}
\def\Ts{T^2}
\def\Us{U^2}
\def\Litwo{\mbox{${\rm{Li}}_{2}$}}
\def\yl{y_l}
\def\yd{y_d}
\newcommand{\ykxyl }{y_{k_1}}
\newcommand{\ykxyll}{y_{k_2}}

\newcommand{\sdkxy }{\sqrt{D_{k}}}

\newcommand{\yLmkl  }{y_{L^*_1}}
\newcommand{\yLmkll }{y_{L^*_2}}

\usepackage{axodraw}

\begin{document}
\begin{center}

{\LARGE\bf $J$ functions for the process ud$\to$WA}

\vspace*{1.5cm}
{\bf D.~Bardin$^{1}$, L.~Kalinovskaya$^{1}$, E.~Uglov$^{1}$, and W. von Schlippe$^{2}$}

\vspace*{\fill}

{\normalsize{\it
$^{1}$ Dzhelepov Laboratory for Nuclear Problems, JINR,        \\
        ul. Joliot-Curie 6, 141980 Dubna, Russia;             \\
$^{2}$ Formerly of PNPI, RAN, Gatchina, 188300, Russia.        \\
}}
\vspace*{\fill}

\end{center}


\vspace*{\fill}

\begin{abstract}
In this paper we present a description of the universal approach for analytic calculations
for a certain class of $J$ functions for six topologies of the boxes
for process $ud\rightarrow WA$.
These functions $J$ arise at the reduction of infrared divergent box diagrams.
The standard Passarino--Veltman reduction of four-point box diagram with an internal photon
line connecting two external lines on the mass shell leads to infrared-divergent
and mass-singular $D_0$ functions.
In the system SANC a systematic procedure is adopted to separate both types of singularities
into the simplest objects, namely $C_0$ functions.
The functions $J$, in turn, are represented as certain linear combinations of the standard
$D_0$ and $C_0$ functions. The subtracted $J$ functions are free of both types of singularities
and are expressed as explicit and compact linear combinations of dilogarithm functions.
We present extensive comparisons of numerical results of SANC with those obtained with the aid
of the LoopTools package.
\end{abstract}

\section{Introduction\label{Introduction}}
The functions $J$ arise in the consideration of infrared divergent box diagrams.
The standard Passarino--Veltman reduction~\cite{Passarino:1978jh} of the four-point box function
with an internal photon line connecting two external lines on the mass shell leads to an
infrared-divergent and mass-singular $D_0$ function.

Functions $J$, in turn, are represented as certain linear combinations of the standard
$D_0$ and $C_0$ functions.
Then the mass singularities are extracted from $J$ to other combinations of $C_0$.
The rest is free of both types of singularities and are expressed as explicit and compact
linear combinations of dilogarithm functions independent of the light fermion masses.
The subtracted $J$ functions, $J_{\rm sub}$, have no mass singularities,
and their compactness leads to stable and very fast calculations.

 $J$ functions arising in the process $f\overline{f}\to AA$
 were originally described in~\cite{Brown:1952eu}.
Later on $J$ functions for four fermion processes  were considered in~\cite{Bardin:1999ak}.
Within the project SANC we propose to introduce infrared finite functions $J$
as a convinient way to disentangle the calculations.

Originally all definitions and steps of calculation for $J$ functions in SANC
were introduced in~\cite{Bardin:2009zz} for the processes $f\overline{f}\to ZZ$,
$f\overline{f}\to ZA$ and $f\overline{f}\to AA$,
Later on we extended our approach by introducing $J$ functions into calculations of
various channnels of the process $udtb\to 0$ at EW NLO level in~\cite{Bardin:2009ix}.
The explicit form the $J_{\rm sub}$ functions depends on the concrete channel of a process,
i.e. we had no universal expression for them.

In this paper we continue the investigation of functions $J$ and $J_{\rm sub}$
arising at the reduction of the infrared divergent box diagrams in the  process $ud \to WA$.
For this process we considered six topologies of boxes with an internal photon line.

We summarize the essential ingredients of our calculation for $J$ functions and point to
the differences with respect to \cite{Bardin:2009zz} and \cite{Bardin:2009ix}.

We provide a universal approach for analytic calculations
of expressions for $J_{\rm uni}$ functions valid for all six topologies for the boxes
of the process $ud\rightarrow WA$.

Section \ref{CalcPro} contains the description of the calculation of this universal function
$J_{\rm uni}$.

In Sections \ref{Topology24}--\ref{Topology66} we give our analytic results for $J_{\rm sub}$
for each box topology.

In Section \ref{Combd0c0} we discuss cancellations of mass singularities in the NLO EW part of
the amplitude of the process under consideration.

In Section \ref{NumRes} we present the numerical comparison for all topologies with results
obtained with the aid of the LoopTools
package~\cite{Hahn:1998yk} for several values of $s$ and $\cos\vartheta$.

In Section 8 we present our conclusions.\\

\newpage

\section{Calculation of $J$ functions for the process $ud \to WA$ \label{CalcPro}}

The calculation of functions $J$ for the process $ud \to WA$  presented here closely
follows the calculation of $J$ for the channel: $ud \to tb$ presented in
the Section 3 of \cite{Bardin:2009ix} and in the earlier paper~\cite{Bardin:2009zz}.

Following Section 14.10 of Ref.\cite{Bardin:1999ak}, the $ff\to bb$ boxes ($f$ stands for a fermion,
$b$ for a boson) could be of seven types which we often call ``topologies'', $T_{1-7}$.

For the process $ud \to WA$, we encountered six infrared divergent box diagrams giving rise to six
$J$ functions, which naturally group into three pairs:\\
1) $T_1,T_3$, Fig.~\ref{topologyT1_3};
2) $T_2,T_4$, Fig.~\ref{topologyT2_4};
3) $T_6,T_{6'}$, Fig.~\ref{topologyT6_6pr}.

The basic definition of a typical function $J$ reads
(see Eq.~(1)--(2) of paper~\cite{Bardin:2009ix}):
\bq
i\pi^2 J=\mu^{4-n}\int d^nq\frac{v(q,p_{i})\cdot v(p_i)}{d_0d_1d_2d_3}\,.
\label{JuniGen}
\eq
The denominators, $d_0,d_1,d_2,d_3$, are the scalar parts of propagators
of virtual particles that a box diagram is comprising; they are inherent to each box topology
under consideration, see Sections~\ref{Topology24}--\ref{Topology66}.
The numerator is the scalar product of two vectors, $v(q,p_{i})$ and $v(p_i)$.
These vectors must satisfy the following two properties.
The first 4-vector
is a linear combination of the integration vector $q$  and of the external 4-momenta $p_{1,2,3,4}$
(ordered counter-clock-wise, see Fig.~\ref{topologyT1_3}, and satisfying the conservation law
$p_1+p_2+p_3+p_4=0$); it is intended to cancel the infrared divergence originating from the
propagator of the virtual photon.
The second 4-vector, another linear combination of external 4-momenta,
must be chosen in a way to simplify the subsequent integration over three Feynman
parameters $z,x,y$, see e.g.~Fig.\ref{topologyT1_3} and the corresponding Eqs.~(\ref{JT1def})
and (\ref{JT3def}) of this paper.

The triple integral over the three Feynman parameters may be expressed by the same
Eqs.(13),~(16)--(17) as given in detail in paper~\cite{Bardin:2009zz}:
\bqa
J = \int \limits_{0}^{1}dx\int\limits_{0}^{1}y\,dy
  N_{xy} \int\limits_{0}^1 dz \frac{z}{(L - z k_{xy}^2)^2}\,,
\label{JAWint3}
\eqa
 with all the variables --- $N_{xy}$, $L$ and the vector squared $k^2_{xy}$ --- being bilinear
forms in Feynman parameters $y,x$ with coefficients made of all
parameters of the problem:
two invariants $P^2_1,\,P^2_2$,
a selection from three $\Qs,\,\Ts,\,\Us$, and all the masses involved.

We recall that we use the standard {\tt SANC} definitions of Mandelstam variables $s,t,u$
\bqa
s=-Q^2=-(p_1+p_2)^2,\qquad t=-T^2=-(p_2+p_3)^2,\qquad u=-U^2=-(p_2+p_4)^2,
\label{stu}
\eqa
where the invariants $\Qs,\,\Ts,\,\Us$ are given in Pauli metrics.

We omit the details of the integrations with respect to $z$ and to $x$ and
 present the integrand of the integration over  $y$.

In Section 3.1 of paper \cite{Bardin:2009ix} we met the case of a function $J$,
when the variables $L$ and $k^2_{xy}$ are linear in $x$ (after one neglects a mass
that does not develop a singularity).
Linearity in $x$ of the vector squared $k^2_{xy}$ and of the variable $L^*=L-k_{xy}^2$
is the key property which makes it possible to introduce one universal function for the calculation
of all six $J$'s which arise in the process $ud \to WA$.

We proceed with the one-dimensional integral, see Eq.(111) of~\cite{Bardin:2009ix}:
\begin{equation}
   J(P^2_1,P^2_2;m_1,m_2,m_3,m_4) = \int\limits_{0}^{1}\,dy I(y),
\label{OneDimInt}
\end{equation}
where we put the entire dummy argument list in the definition of $J$.

For the integrand $I(y)$ one obtains (see Eq.(112))\footnote{There is a misprint in the last term
of~Eq.(112) of~Ref.\cite{Bardin:2009ix};
the correct one is the last term of Eq.(\ref{Integrand})  of this paper.}
of~\cite{Bardin:2009ix}:
\begin{eqnarray}
I(y)=\left( -\frac{1}{k^2_{{xy}|y}} - \frac{1}{T^2_y-i\epsilon} \right)
             \left[\ln(L^*{_|{_{_y}}})-\ln(P^* (1-y))\right],
\label{Integrand}
\end{eqnarray}
with ingredients of~Eq.(113) of~Ref.\cite{Bardin:2009ix}:
\begin{eqnarray}
\label{Ingr}
P^* &=& \Pl+m_{3}^2-i\epsilon\,,
\\
T^2_y       &=&\Pll y (1-y)+m_{1}^2y+m_{4}^2(1-y)\,,
\label{Ty}
\\
k^2_{{xy}|y} &=& m_{2}^2 y(1-y)-m_{1}^2y+\Pl(1-y)\,,
\label{kxy}
\\
L^*{_|{_{_y}}}&=&-m_{2}^2 y(1-y)+m_{1}^2y+m_{3}^2(1-y)-i\epsilon\,.
\label{Lstar}
\end{eqnarray}
Here, as previously,
$L^*{_|{_{_y}}}=L^*(x = y,\,y)$ 
and
$k_{xy}^2{_|{_{_y}}}=k_{xy}^2(x = y,\,y)$.
The differences are: change of notation $P^2\to P^*$ and
the use here of dummy invariants $P^2_{1,2}$ instead of the physical ones $\Qs,\,\Ts$
which were used in \cite{Bardin:2009ix} for a specific process.

Two sets of topologies arise in our investigation.
The first one
is $T_2, T_4$ and $T_6, T_6^{'}$.
For this set we received the universal answer, $J_{\rm uni}$, for the integral (\ref{OneDimInt})
in terms of four calls to the auxiliary function of three arguments.
The second set consists of two topologies $T_1, T_3$.
The answer for this case, $J^0_{\rm uni}$, is the limit of the previous one.
It is simpler and can be expressed via three calls to a simpler
auxiliary function of two arguments.

For all topologies we take the limit of vanishing light quark masses.
The mass of the quark which is not coupled to the photon may be set equal to zero,
while that for the quark coupled to the photon develops a mass singular logarithm.
We keep quark masses in arguments of logarithmic functions and neglect them everywhere else.

\subsection{Result of integration over $y$ of the first set of topologies}

One can get a universal result of the integration, $J_{\rm uni}$, for the first set of topologies
which is expressed in terms of the auxiliary function ${\cal{L}}(a,b,c)$:

\def\sdTy{\sqrt{D_T}}
\def\yTyl{y_{T_1}}
\def\yTyll{y_{T_2}}
\bqa
 J_{\rm uni}(P^2_1,P^2_2;m_1,m_2,m_3,m_4)
 &=& -\frac{1}{\sdkxy}\left({\cal{L}}(\yLmkl,\yLmkll,\ykxyl)-{\cal{L}}(\yLmkl,\yLmkll,\ykxyll)\right)\nll
 &&  -\frac{1}{\sdTy} \left({\cal{L}}(\yLmkl,\yLmkll,\yTyl) -{\cal{L}}(\yLmkl,\yLmkll,\yTyll) \right).
\label{AnswerFirstSet}
\eqa

The auxiliary function depends on three arguments:
\bqa
 {\cal{L}}(a,b,c)=  {\cal{M}}(a,c) +  {\cal{M}}(b,c) -  {\cal{M}}(1,c)
       -\ln\left(1-\frac{1}{c}\right) \ln\left(\frac{P^*}{m_3^2}\right),
\label{FirstAuxiliary}
\eqa
with the ``master integral'':
\bqa
{\cal{M}}(\yd,\yl) &=& \int^1_0 \frac{dy}{(y-\yd)\ln(1-y/\yl)} \nll
&=&
            \ln\left(1-\frac{\yd}{\yl}\right)
            \ln\left(1-\frac{1  }{\yd}\right)
           -\Litwo\left(\frac{1-\yd}{\yl-\yd}\right)
           +\Litwo\left(\frac{ -\yd}{\yl-\yd}\right).
\label{Master}
\eqa

The arguments of the auxiliary functions in (\ref{AnswerFirstSet}) are the
roots of quadratic trinomials:

$\bullet$ Roots of the quadratic trinomial~(\ref{Ty}):
\begin{equation}
      y_{{T}_{1,2}} = \frac{-b_{T} \pm \sqrt{D_{T}}}{(-2 P^2_2)},
~~\mbox{where} ~~ b_{T} = - m_4^2 + m_1^2 + P^2_2, ~ 
      D_{T} = b_{T}^2 + 4 P^2_2 (m_4^2-i\varepsilon).
\label{RootsTy}
\end{equation}

$\bullet$ Roots of the quadratic trinomial~(\ref{kxy}):
\begin{equation}
      y_{{k}_{1,2}}=\frac{-b_{k} \pm \sqrt{D_{k}}}{(-2 m_2^2)},
~~\mbox{where} ~~  b_{k} = - m_1^2 + m_2^2 - P^2_1,
  ~  D_{k}= b^2_{k}+4  m_2^2 (P^2_1+i\varepsilon).
\end{equation}

$\bullet$ Roots of the quadratic trinomial~(\ref{Lstar}):
\begin{equation}
      y_{L^*_{1,2}}=\frac{-b_{L} \pm \sqrt{D_{L}}}{2 m_2^2},
~~\mbox{where} ~~ b_{L} = - m_3^2 + m_1^2 - m_2^2,
   ~   D_{L}=b_{L}^2-4 m_2^2 (m_3^2-i\varepsilon).
\end{equation}

\subsection{Result of integration of the second set of topologies}

This result is a particular case of the previous one~(\ref{AnswerFirstSet}) at $m_1=m_3$ and
$m_2=0$; it reads:
\bqa
J^0_{\rm uni}(P^2_1,P^2_2,m_3,0,m_3,m_4) =
             -\frac{1}{P^*}\Biggl[{\cal{L}}_0\left(\frac{\Qs}{P^*}\right)\Biggr]
             +\frac{1}{\sdTy} \Biggl[{\cal{L}}_0(\yTyl) - {\cal{L}}_0(\yTyll)\Biggr].
\label{AnswerSecondSet}
\eqa
The ingredients (\ref{Ingr})--(\ref{Lstar}) simplify considerably and in
this case the auxiliary function reduces to the function of one variable ${\cal{L}}_0(a)$:
\bqa
 {\cal{L}}_0(a)=  {\cal{M}}(1,a)
       +\ln\left(1-\frac{1}{a}\right) \ln\left(\frac{P^*}{m_3^2}\right),
\label{SecondAuxiliary}
\eqa
where $P^*$ is given by Eq.~(\ref{Ingr}), $\yTyl,\,\yTyll$ are roots (\ref{RootsTy}) of the quadratic
trinomial~(\ref{Ty}) and the master integral ${\cal{M}}(\yd,\yl)$ is the same as before, Eq.~(\ref{Master}).



\newpage

\section{Topologies $T_2,T_4$\label{Topology24}}

\subsection{Definition of functions $J^{T_2,T_4}$}
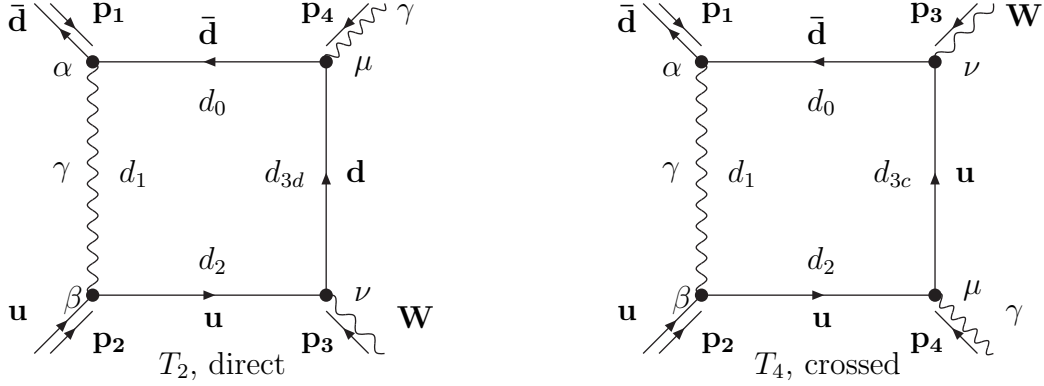
\begin{figure}[!h]
\[
\begin{array}{ccc}
  \vcenter{\hbox{
\begin{picture}(132,132)(0,0)
 \Text(-32,132)[lt]{$\bf \bar d$}
 \Text(-32,18)[lt]{$\bf      u$}
 \Text(115,132)[lt]{$\bf \gamma$}
 \Text(115,18)[lt]{$\bf     W$}

 \ArrowLine(88,22)(88,110)
 \ArrowLine(88,110)(0,110)
 \ArrowLine(0,22)(88,22)

 \Vertex(88,22){2.5}
 \Photon(88,22)(110,0){2}{3}
 \Photon(110,132)(88,110){2}{5}

 \Vertex(0,110){2.5}
 \ArrowLine(0,110)(-22,132)

 \Photon(0,110)(0,22){2}{11}
 \Vertex(0,22){2.5}

 \ArrowLine(-22,0)(0,22)
 \ArrowLine(-16,132)(0,116)
 \ArrowLine(-16,0)(0,16)

 \ArrowLine(104,132)(88,116)
 \Vertex(88,110){2.5}
 \ArrowLine(104,0)(88,16)

 \Text(-15,105)[lb]{$\bf \alpha$}
 \Text(95,105)[lb]{ $\bf \mu$}
 \Text(95,20)[lb]{  $\bf \nu$}
 \Text(-15,15)[lb]{ $\bf \beta $}
 \Text(-4,125)[lb]{ $\bf p_1$}
 \Text(-4,0)[lb]{   $\bf p_2$}
 \Text(75,125)[lb]{ $\bf p_4$}
 \Text(75,0 )[lb]{  $\bf p_3$}

  \Text(-15,65)[lb]{$\bf \gamma$}
  \Text(40,117)[lb]{$\bf \bar d$}
  \Text(92,65)[lb]{ $\bf      d$}
  \Text(38,10)[lb]{ $\bf u$}

  \Text(40,90)[lb]{$d_0$}
  \Text(10,63)[lb]{$d_1$}
  \Text(65,63)[lb]{$d_{3d}$}
  \Text(40,30)[lb]{$d_2$}

 \Text(25,-10)[lb]{$T_2$, direct}
\end{picture}
}}
& \qquad \qquad \qquad &
  \vcenter{\hbox{
\begin{picture}(132,132)(0,0)
 \ArrowLine(88,22)(88,110)
 \Vertex(88,22){2.5}
 \Photon(88,22)(110,0){2}{5}
 \ArrowLine(88,110)(0,110)
 \Vertex(0,110){2.5}
 \ArrowLine(0,110)(-22,132)
 \Photon(0,110)(0,22){2}{11}
 \Vertex(0,22){2.5}
 \ArrowLine(-22,0)(0,22)
 \ArrowLine(0,22)(88,22)
 \ArrowLine(104,132)(88,116)
 \Vertex(88,110){2.5}
 \Photon(110,132)(88,110){2}{3}
 \ArrowLine(104,0)(88,16)
 \ArrowLine(-16,0)(0,16)
 \ArrowLine(-16,132)(0,116)
 \Text(-32,132)[lt]{$\bf \bar d$}

 \Text(-32,18)[lt]{$\bf u$}
 \Text(115,132)[lt]{$\bf W $}
 \Text(115,18)[lt]{$\bf   \gamma$}
 \Text(-15,105)[lb]{$\bf \alpha$}
 \Text(95,105)[lb]{ $\bf \nu$}
 \Text(95,20)[lb]{  $\bf \mu$}
 \Text(-15,15)[lb]{ $\bf \beta$}
 \Text(-4,125)[lb]{ $\bf p_1$}
 \Text(-4,0)[lb]{   $\bf p_2$}
 \Text(75,125)[lb]{ $\bf p_3$}
 \Text(75,0 )[lb]{  $\bf p_4$}

  \Text(-15,65)[lb]{$\bf \gamma$}
  \Text(40,117)[lb]{$\bf \bar d$}
  \Text(92,65)[lb]{ $\bf u$}
  \Text(38,10)[lb]{ $\bf u$}

  \Text(40,90)[lb]{$d_0$}
  \Text(10,63)[lb]{$d_1$}
  \Text(65,63)[lb]{$d_{3c}$}
  \Text(40,30)[lb]{$d_2$}

 \Text(20,-10)[lb]{$T_4$, crossed}
\end{picture}}}
\end{array}
\]
\caption{Process $u\bar{d} \to WA$. Box topologies $T_2$ and $T_4$.}
\label{topologyT2_4}
\end{figure}

 For the process $u {\bar d} \rightarrow WA$,
 the box diagrams for the topologies $T_2,T_4$ are shown in Fig.~\ref{topologyT2_4}.
 They are of the {\it direct} and {\it crossed} type, respectively.

\vspace*{-3mm}

The direct channel function $J^{T_2}$ is defined by the following equation:
\bqa
i\pi^2 J^{T_2}
=\mu^{4-n}\int d^n q~ \frac{\ds 2(q+p_1) p_4}{\ds d_0(\mdn) d_1(0) d_2(\mup) d_{3d}(\mdn)};
\label{JT2def}
\eqa
its arguments are not shown on purpose.
The relevant masses enter through the scalar denominators $d_i$:
\bqa
d_0 &=&q^2+\mdn^2\\
d_1 &=&(q+p_1)^2\\
d_2 &=&(q+p_1+p_2)^2+\mup^2\\
d_{3d}&=&(q-p_4)^2+\mdn^2\\
d_{3c}&=&(q-p_3)^2+\mup^2
\eqa

The direct function $J^{T_2}$ is expressed via the universal function

\vspace*{-5mm}

\bqa
J_{\rm uni}(P^2_1,P^2_2;m_1,m_2,m_3,m_4),
\label{Junideclar}
\eqa
\vspace*{-5mm}

\noindent given by Eq.~(\ref{AnswerFirstSet}) of the previous Section:

\vspace*{-5mm}

\bqa
J^{T_2}=J_{\rm uni}(T^2,Q^2;\mup,\mw,\mdn,\mdn).
\label{JuniT2}
\eqa
For the related cross channel topology the definition of $J^{T_4}$ looks similarly:
\bqa
i\pi^2 J^{T_4}
   =\mu^{4-n}\int d^n q~ \frac{\ds -2(q+p_1)p_4}{\ds d_0(\mdn) d_1(0) d_2(\mup) d_{3c}(\mup)}.
\label{JT4def}
\eqa
The same comment about its arguments is also valid here, and in terms of $J_{\rm uni}$
one gets
\bqa
J^{T_4}=J_{\rm uni}(U^2,Q^2;\mdn,\mw,\mup,\mup).
\label{JuniT4}
\eqa


{$\bullet$ $J^{T_2,T_4}$ as functions of $D_0$ and $C_0$}

 For topology $T_2$ using the standard Passarino--Veltman reduction it is possible
to derive relations between infrared- and mass-singular functions
$$D_0(-\mdn^2,-\mup^2,-\mw^2,0,\Qs,\Ts;\mdn,0,\mup,\mdn)
\;\mbox{and}\;
C_0(-\mdn^2,-\mup^2,\Qs;\mdn,0,\mup)$$
and our infrared finite but mass-singular $J$-function under consideration,
$J^{T_2}$ ,
and another $C_0(-\mup^2,-\mw^2,\Ts;0,\mup,\mdn)$ with mass singularity.

This relation, exact in all masses, is
\bqa
J^{T_2} &=&
 \left(\Ts+\mdn^2\right) D_0(-\mdn^2,-\mup^2,-\mw^2,0,\Qs,\Ts;\mdn,0,\mup,\mdn)
\nll &&
      - C_0(-\mdn^2,-\mup^2,\Qs;\mdn,0,\mup)
      + C_0(-\mup^2,-\mw^2,\Ts;0,\mup,\mdn).
\label{relT2}
\eqa
For $J^{T_4}$, a similar relation holds. However, neglecting terms proportional
to the quark mass powers $m^{2}_{u,d}/Q^{2}$, one gets
\bqa
J^{T_4} &=&
\left(\Us+\mup^2\right) D_0(-\mdn^2,-\mup^2,0,-\mw^2,\Qs,\Us;\mdn,0,\mup,\mup)
\nll  &&
         - C_0(-\mdn^2,-\mup^2,\Qs;\mdn,0,\mup)
         + C_0(-\mdn^2,-\mw^2,\Us;0,\mdn,\mup).
\label{relT4}
\eqa
It is a typical property of such relations to be exact in masses for {\it direct}
boxes, but for {\it crossed} boxes only up to some mass power terms, which we do not control anyway.

The great advantage of relations~(\ref{relT2})--(\ref{relT4}) 
is the following. The complicated object $D_0$,
containing an infrared divergence, is excluded in favor of explicitly computed
functions $J^{T_2}$ and $J^{T_4}$
and the simplest infrared-divergent object $C_0(-\mdn^2,-\mup^2,\Qs;\mdn,0,\mup)$,
whose infrared divergences can be regularized by any method:
by a photon mass, by dimensional regularization or by the width of an unstable particle.
Examples of $C_0$ functions regularized by the width can be found in
Ref.~\cite{Bardin:2009wv}.\\

\newpage

$\bullet$ Subtracted functions $J^{T_{2,4}}_{\rm{sub}}$

Adding to the relations~(\ref{relT2})--(\ref{relT4}) the other pinches of the primary $D_0$
(which in general are mass-singular) with correspondingly adjusted kinematical coefficients,
one gets the ``subtracted functions'' $J^{T_{2,4}}_{\rm{sub}}$
which are free of quark mass singularities:
\bqa
J^{T_{2}}_{\rm{sub}} &=&
\left(\Ts+\mdn^2\right) D_0(-\mdn^2,-\mup^2,-\mw^2,0,\Qs,\Ts;\mdn,0,\mup,\mdn)
\nll &&
      - C_0(-\mdn^2,-\mup^2,\Qs;\mdn,0,\mup)
      -\frac{\Ts}{\Qs} C_0(0,-\mdn^2,\Ts;\mdn,\mdn,0)
\nll &&
      -\frac{\Ts+\mw^2}{\Qs} C_0(-\mup^2,-\mw^2,\Ts;0,\mup,\mdn)
\nll &&
      -\frac{\Qs+\mw^2}{\Qs} C_0(0,      -\mw^2,\Qs;\mdn,\mdn,\mup),
\label{answT2sub}
\eqa
and
\bqa
J^{T_{4}}_{\rm{sub}} &=&
\left(\Us+\mup^2\right) D_0(-\mdn^2,-\mup^2,0,-\mw^2,\Qs,\Us;\mdn,0,\mup,\mup)
\nll  &&
       - C_0(-\mdn^2,-\mup^2,\Qs;\mdn,0,\mup)
       -\frac{\Us}{\Qs}  C_0(0,-\mup^2,\Us;\mup,\mup,0)  
\nll  &&
       -\frac{\Us+\mw^2}{\Qs} C_0(-\mdn^2,-\mw^2,\Us;0,\mdn,\mup) 
\nll  &&
       -\frac{\Qs+\mw^2}{\Qs} C_0(0,-\mw^2,\Qs;\mup,\mup,\mdn).   
\label{answT4sub}
\eqa

\newpage

\subsection{Pinches of topologies $T_2$ and $T_4$}
Each box diagram contains four three-point pinches. Here we present pinch diagrams and their
expressions in terms of the corresponding $C_0$ functions for box topologies $T_2$ and $T_4$.
Note that all four pinches contribute to the relations~(\ref{answT2sub})--(\ref{answT4sub}).
Furthermore, we give the explicit expressions for three infrared-finite and mass-singular pinches
$C_{0,1-3}^{T_{2}}$ in the limit $m_u=m_d=0$, i.e. keeping these masses only in arguments of logarithmic
functions.

\vspace*{.5cm}

$\bullet$ {Topology $T_2$ pinches}

\vspace*{1.5cm}

\begin{figure}[!h]
\[
\begin{array}{cccc}
  \vcenter{\hbox{
\begin{picture}(88,88)(0,0)
\ArrowLine(0,66)(44,88)
\ArrowLine(44,88)(0,110)
\Photon(0,66)(0,110){2}{7}
\Photon(44,88)(66,66){2}{3}
\Photon(44,88)(66,110){2}{5}

\ArrowLine(-22,44)(0,66)
\ArrowLine(0,110)(-22,132)

\Text(-25,58)[lb]{ $\bf u$}
\Text(-25,115)[lb]{$\bf \bar d$}
\Text(22,60)[lb]{  $\bf u$}
\Text(22,110)[lb]{ $\bf \bar d$}
\Text(60,55)[lb]{  $\bf W$}
\Text(-20,82)[lb]{ $\bf \gamma$}
\Text(60,115)[lb]{ $\bf \gamma$}
\Text(22,30)[lb]{$C_{0,\rm{IRD}}^{T_2}$}
\end{picture}}}&
  \vcenter{\hbox{
\begin{picture}(88,88)(0,0)
\ArrowLine(0,110)(-22,132)
\ArrowLine(44,110)(0,110)
\ArrowLine(22,66)(44,110)
\Photon(22,66)(0,110){2}{9}
\Photon(44,110)(66,132){2}{5}
\ArrowLine(0,44)(22,66)
\Photon(22,66)(44,44){2}{3}
\Text(-10,125)[lb]{$\bf \bar d$}
\Text(12,117)[lb]{ $\bf \bar d$}
\Text(44,125)[lb]{ $\bf \gamma$}
\Text(40,88)[lb]{  $\bf      d$}
\Text(-7,84)[lb]{  $\bf \gamma$}
\Text(-5,55)[lb]{   $\bf u$}
\Text(40,55)[lb]{  $\bf W$}
\Text(12,30)[lb]{$C_{0,1}^{T_2}$}
\end{picture}}}
&
  \vcenter{\hbox{
\begin{picture}(88,88)(0,0)
\ArrowLine(22,110)(0,132)
\Photon(0,66)(22,110){2}{9}
\ArrowLine(44,66)(22,110)
\ArrowLine(0,66)(44,66)
\Photon(22,110)(44,132){2}{5}
\ArrowLine(-22,44)(0,66)
\Photon(44,66)(66,44){2}{3}
\Text(-10,125)[lb]{$\bf \bar d$}
\Text(15,72)[lb]{  $\bf u$}
\Text(48,125)[lb]{ $\bf \gamma$}
\Text(40,88)[lb]{  $\bf      d$}
\Text(-7,84)[lb]{  $\bf \gamma$}
\Text(-26,55)[lb]{ $\bf u$}
\Text(60,55)[lb]{  $\bf W$}
\Text(12,30)[lb]{$C_{0,2}^{T_2}$}
\end{picture}}}
&
  \vcenter{\hbox{
\begin{picture}(88,88)(-10,0)
\ArrowLine(-22,66)(0,88)
\ArrowLine(0,88)(-22,110)
\ArrowLine(44,110)(0,88)
\ArrowLine(0,88)(44,66)
\ArrowLine(44,66)(44,110)
\Photon(44,110)(66,132){2}{5}
\Photon(44,66)(66,44){2}{3}
\Text(-10,60)[lb]{ $\bf u$}
\Text(-10,110)[lb]{$\bf \bar d$}
\Text(15,60)[lb]{  $\bf u$}
\Text(15,110)[lb]{ $\bf \bar d$}
\Text(60,115)[lb]{ $\bf \gamma$}
\Text(50,88)[lb]{  $\bf      d$}
\Text(60,55)[lb]{  $\bf W$}
\Text(12,30)[lb]{$C_{0,3}^{T_2}$}
\end{picture}}}
\end{array}
\]

\vspace*{-1cm}

\caption{Diagrams of pinches for box topology ~$T_2$.}
\label{topologyTIIpinch}
\end{figure}
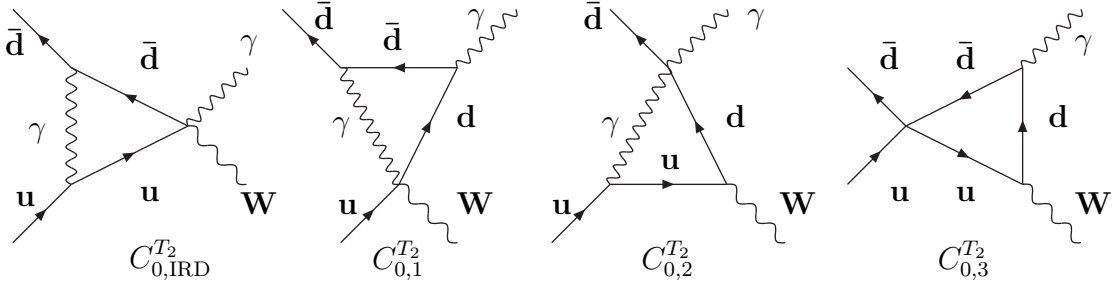
\vspace*{-.5cm}

\bqa
C_{0,\rm{IRD}}^{T_2}  &=& C_0(-\mdn^2,-\mup^2,\Qs;\mdn,0,\mup),
\nll
C_{0,1}^{T_2}   &=& C_0(0,-\mdn^2,\Ts;\mdn,\mdn,0)\;=\;
\frac{1}{\Ts}\left[\frac{1}{2}
\ln^2\left(\frac{\Ts-\ieps}{\mdn^2}\right)+2\zeta(2)\right],
\nll
C_{0,2}^{T_2}   &=& C_0(-\mup^2,-\mw^2,\Ts;0,\mup,\mdn)
\nll &=&
      \frac{1}{\Ts+\mw^2}\Biggl\{
  \ln\left[-(1-\ieps)\right] \ln\left[\frac{(\Ts+\mw^2)^2 \mw^2}{(\Ts)^3}\right]
\nll &&
    +\Litwo\left(1+\frac{\mw^2-\ieps}{\Ts}\right)
 -\Litwo\left( -\frac{\mw^2-\ieps}{\Ts}\right)
    +4 \zeta(2)
\nll &&
 +\ln\left[-(1-\ieps)\frac{\Ts}{\mw^2}\right] \ln\left(\frac{\Ts+\mw^2}{\mup^2}\right)
           -\frac{1}{2}\ln^2\left(\frac{\Ts}{\mw^2}\right)
                           \Biggr\},
\nll
C_{0,3}^{T_2}   &=& C_0(0,-\mw^2,\Qs;\mdn,\mdn,\mup)
\nll &=&
-\frac{1}{\Qs+\mw^2}
\Biggl\{
\ln\left(-\frac{\mw^2}{\Qs}\right)\ln\left(-\frac{\Qs}{\mdn^2}\right)
\nll &&
+\frac{1}{2}\ln\left(-\frac{\mw^2}{\Qs}\right)
\Biggl[\ln\left(\frac{\mw^2+\ieps}{\Qs}\right)-\ln(-1+\ieps)\Biggr]
\Biggr\}.
\label{pinchesT2}
\eqa
As is seen, for $C_{0,2}^{T_2}$ and $C_{0,3}^{T_2}$ only one quark mass ($\mup$ and $\mdn$, respectively)
appears on the right-hand-side of the resulting expession. This means that the singularity over the
other mass does not develop and may be safely neglected.

\vspace*{.5cm}

$\bullet$ {Topology $T_4$ pinches}\\

\vspace*{1cm}

\begin{figure}[!h]
\[
\begin{array}{ccccc}
  \vcenter{\hbox{
\begin{picture}(88,88)(0,0)
\ArrowLine(0,66)(44,88)
\ArrowLine(44,88)(0,110)
\Photon(0,66)(0,110){2}{7}
\Photon(44,88)(66,66){2}{5}
\Photon(44,88)(66,110){2}{3}

\ArrowLine(-22,44)(0,66)
\ArrowLine(0,110)(-22,132)

\Text(-25,58)[lb]{ $\bf u$}
\Text(-25,115)[lb]{$\bf \bar d$}
\Text(22,60)[lb]{  $\bf u$}
\Text(22,110)[lb]{ $\bf \bar d$}
\Text(60,55)[lb]{  $\bf \gamma$}
\Text(-20,82)[lb]{ $\bf \gamma$}
\Text(60,115)[lb]{ $\bf W$}
\Text(22,30)[lb]{$C_{0,\rm{IRD}}^{T_4}$}
\end{picture}}}
&
  \vcenter{\hbox{
\begin{picture}(88,88)(0,0)
\ArrowLine(22,110)(0,132)
\Photon(22,110)(44,132){2}{3}
\Photon(0,66)(22,110){2}{7}
\Photon(44,66)(66,44){2}{5}
\ArrowLine(44,66)(22,110)
\ArrowLine(0,66)(44,66)
\ArrowLine(-22,44)(0,66)
\Text(-10,125)[lb]{$\bf \bar d$}
\Text( 17,55)[lb]{ $\bf u$}
\Text(48,125)[lb]{ $\bf W$}
\Text(40,88)[lb]{  $\bf u$}
\Text(-5,84)[lb]{  $\bf \gamma$}
\Text(-27,55)[lb]{ $\bf u$}
\Text(58,55)[lb]{  $\bf \gamma$}
\Text(12,30)[lb]{$C_{0,1}^{T_4}$}
\end{picture}}}
&
  \vcenter{\hbox{
\begin{picture}(88,88)(0,0)
\ArrowLine(0,110)(-22,132)
\ArrowLine(44,110)(0,110)
\ArrowLine(22,66)(44,110)
\ArrowLine(0,44)(22,66)
\Photon(22,66)(44,44){2}{5}
\Photon(22,66)(0,110){2}{7}
\Photon(44,110)(66,132){2}{3}
\Text(-10,125)[lb]{$\bf \bar d$}
\Text(17,117)[lb]{ $\bf \bar d$}
\Text(42,125)[lb]{ $\bf W$}
\Text(40,88)[lb]{  $\bf u$}
\Text(-7,84)[lb]{  $\bf \gamma$}
\Text(-5,55)[lb]{  $\bf u$}
\Text(38,55)[lb]{  $\bf \gamma$}
\Text(12,30)[lb]{$C_{0,2}^{T_4}$}
\end{picture}}}
&
  \vcenter{\hbox{
\begin{picture}(88,88)(0,0)
\ArrowLine(-22,66)(0,88)
\ArrowLine(0,88)(-22,110)
\ArrowLine(44,110)(0,88)
\ArrowLine(0,88)(44,66)
\ArrowLine(44,66)(44,110)
\Photon(44,110)(66,132){2}{3}
\Photon(44,66)(66,44){2}{5}
\Text(-15,60)[lb]{ $\bf u$}
\Text(-15,110)[lb]{$\bf \bar d$}
\Text(22,60)[lb]{  $\bf u$}
\Text(22,110)[lb]{ $\bf \bar d$}
\Text(60,55)[lb]{  $\bf \gamma$}
\Text(50,88)[lb]{  $\bf u$}
\Text(60,115)[lb]{ $\bf W$}
\Text(12,30)[lb]{$C_{0,3}^{T_4}$}
\end{picture}}}
\end{array}
\]

\vspace*{-1.5cm}

\caption{Diagrams of pinches for box topology ~$T_4$.}
\label{topologyTIVpinch}
\end{figure}
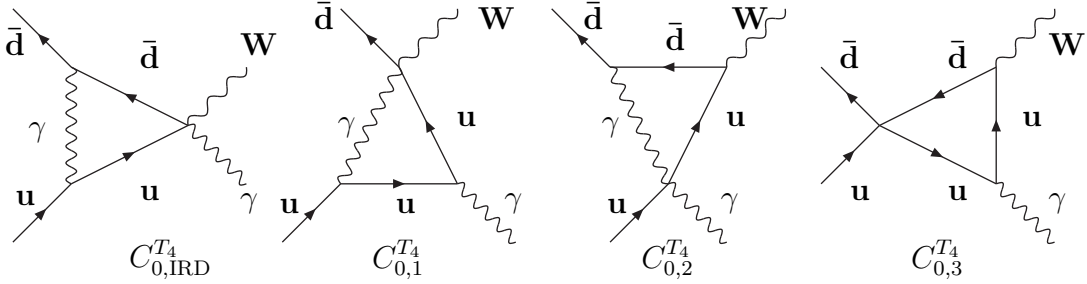

The pinches for box topology $T_4$ in terms of $C_0$ functions are:
\bqa
C_{0,\rm{IRD}}^{T_4} &=& C_0(-\mdn^2,-\mup^2,\Qs;\mdn,0,\mup),
\nll
C_{0,1}^{T_4}   &=& C_0(0,-\mup^2,\Us;\mup,\mup,0),
\nll
C_{0,2}^{T_4}   &=& C_0(-\mdn^2,-\mw^2,\Us;0,\mdn,\mup),
\nll
C_{0,3}^{T_4}   &=& C_0(0,-\mw^2,\Qs;\mup,\mup,\mdn).
\label{pinchesT4}
\eqa
Three pinches $C_{0,1-3}^{T_{4}}$ are obtained from $C_{0,1-3}^{T_{2}}$ by the replacements $\Ts\to \Us$ and
$\mdn\leftrightarrow\mup$.

\subsection{The final manipulations with functions $J^{T_2,T_4}_{\rm{sub}}$\label{finalmanT24}}
We exploit Eqs.(\ref{answT2sub})--(\ref{answT4sub}), as well as similar expressions for the other box
topologies below, in two ways. Let us exemplify this with Eqs.(\ref{relT2}),~(\ref{answT2sub}) for the
topology $T_2$.

In the first way, we can combine the latter equations to exclude infrared divergent $D_0$ and $C_0$
and use notations (\ref{pinchesT2}) for pinches:
\bqa
J^{T_{2}}_{\rm{sub}} &=& J^{T_2} -\frac{\Ts}{\Qs}C_{0,1}^{T_2}
      -\left(1+\frac{\Ts+\mw^2}{\Qs}\right) C_{0,2}^{T_2}
      -\frac{\Qs+\mw^2}{\Qs}C_{0,3}^{T_2}.
\label{answT2expl}
\eqa

As the next step in deriving the function $J^{T_2}_{\rm{sub}}$ we substitute $J^{T_2}$ via
Eq.(\ref{JuniT2}) and pinches via expressions explicitly presented in the previous Section,
Eqs.(\ref{pinchesT2}).
Then the limit in the masses $\mdn\to 0$ and $\mup\to 0$ is calculated. The final answer, expressed
in terms of dilogarithms, does not contain logarithmic mass singularities and is very compact:
\bqa
J^{T_2,T_4}_{\rm{sub}}(\Qs,\Ps,\mw^2) &=&
\frac{1}{\Qs}\Biggl[
            -\ln^2  \left(-\frac{\mw^2}{\Qs+\mw^2}\right)+\ln^2\left(\frac{\mw^2}{\Ps}\right)
\nll  &&
            +2\ln   \left(-\frac{\mw^2+\ieps}{\Ps}\right)
              \ln   \left(-\frac{\Ps+\mw^2  }{\Qs}\right)
\nll  &&
            -2\Litwo\left( \frac{\mw^2}{\Qs+\mw^2+\ieps}\right)
            +2\Litwo\left(-\frac{\mw^2}{\Ps+\ieps}      \right) \Biggr].
\label{answT2T4sim}
\eqa

Here we present the final answers for $J^{T_2,T_4}_{\rm{sub}}$
for both topologies $T_{2,4}$ restoring the list of physical arguments:
$\Ps=\Ts$ for topology $T_2$, and $\Ps=\Us$ for topology $T_4$.
(Note also, that for Eq.~(\ref{answT2T4sim}) we limit ourselves to the case
of $s$-channel kinematics with $\Qs < 0$ and $\Ps > 0$.)

In the second way, we may invert Eq.(\ref{answT2sub}) to
exclude the infrared divergent $D_0$ in favour of $J^{T_2}_{\rm{sub}}$ and for $C_{0,i}^{T_2}$ pinches, $i=1,2,3$.
$C_{0,\rm{IRD}}^{T_2}$ is assigned to the QED part of the NLO EW correction, (\ref{invertedT2subQED}),
whereas $J^{T_2}_{\rm{sub}}$ and the three pinches $C_{0,1-3}^{T_{2}}$
are assigned to the PW (Pure Weak) part of the NLO EW correction, (\ref{invertedT2subEW}):
\bqa
&&\left(\Ts+\mdn^2\right) D_0(-\mdn^2,-\mup^2,-\mw^2,0,\Qs,\Ts;\mdn,0,\mup,\mdn) =
\nll &&
      + C_0(-\mdn^2,-\mup^2,\Qs;\mdn,0,\mup)
\label{invertedT2subQED}
\\ &&
      - J^{T_{2}}_{\rm{sub}}(\Qs,\Ps,\mw^2)
      +\frac{\Ts}{\Qs} C_0(0,-\mdn^2,\Ts;\mdn,\mdn,0)
\label{invertedT2subEW}
\\ &&
      +\frac{\Ts+\mw^2}{\Qs} C_0(-\mup^2,-\mw^2,\Ts;0,\mup,\mdn)
      +\frac{\Qs+\mw^2}{\Qs} C_0(0,-\mw^2,\Qs;\mdn,\mdn,\mup).
\nonumber
\eqa

For the box topology $T_4$, its functions are treated in the same way.

\newpage

\section{Topologies $T_1$, $T_3$\label{Topology13}}

\subsection{Definition of functions $J^{T_1,T_3}$}
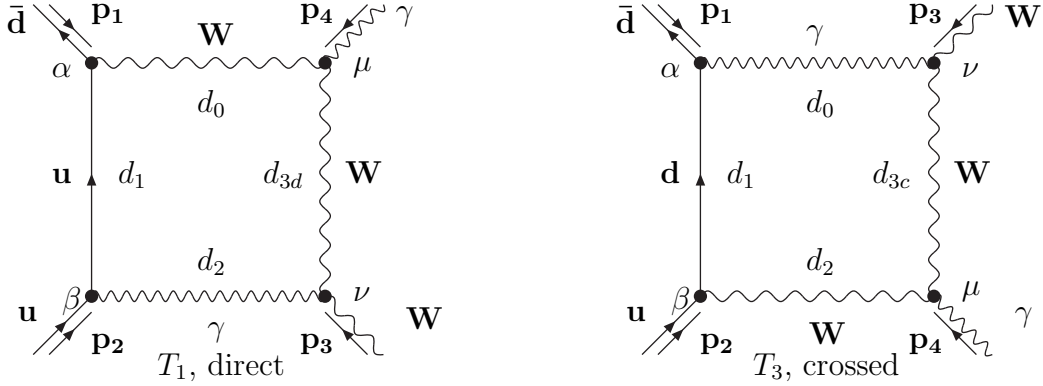
\begin{figure}[!h]
\[
\begin{array}{ccc}
  \vcenter{\hbox{
\begin{picture}(132,132)(0,0)
 \Vertex(88,22){2.5}
 \Photon(88,22)(110,0){2}{3}
 \Photon(110,132)(88,110){2}{5}
 \Vertex(0,110){2.5}
 \ArrowLine(0,110)(-22,132)
 \ArrowLine(0,22)(0,110)
 \Vertex(0,22){2.5}
 \ArrowLine(-22,0)(0,22)
 \Photon(88,110)(0,110){2}{8}
 \Photon(0,22)(88,22){2}{15}
 \Photon(88,22)(88,110){2}{8}
 \ArrowLine(104,132)(88,116)
 \Vertex(88,110){2.5}
 \ArrowLine(104,0)(88,16)
 \ArrowLine(-16,0)(0,16)
 \ArrowLine(-16,132)(0,116)

 \Text(-32,132)[lt]{$\bf \bar d$}
 \Text(-32,18)[lt]{ $\bf      u$}
 \Text(115,132)[lt]{$\bf \gamma$}
 \Text(115,18)[lt]{ $\bf     W$}

 \Text(-15,105)[lb]{$\bf \alpha$}
 \Text(95,105)[lb]{ $\bf \mu$}
 \Text(95,20)[lb]{  $\bf \nu$}
 \Text(-15,15)[lb]{ $\bf \beta $}

 \Text(-4,125)[lb]{ $\bf p_1$}
 \Text(-4,0)[lb]{   $\bf p_2$}
 \Text(75,125)[lb]{ $\bf p_4$}
 \Text(75,0 )[lb]{  $\bf p_3$}

  \Text(-15,65)[lb]{$\bf u$}
  \Text(40,117)[lb]{$\bf W$}
  \Text(92,65)[lb]{ $\bf W$}
  \Text(40,12)[lt]{ $\bf \gamma$}

  \Text(40,90)[lb]{$d_0$}
  \Text(10,63)[lb]{$d_1$}
  \Text(65,63)[lb]{$d_{3d}$}
  \Text(40,30)[lb]{$d_2$}

  \Text(25,-10)[lb]{$T_1$, direct}
\end{picture}
}}
& \qquad \qquad \qquad &
  \vcenter{\hbox{
\begin{picture}(132,132)(0,0)
 \Vertex(88,22){2.5}

 \Vertex(0,110){2.5}
 \ArrowLine(0,110)(-22,132)
 \ArrowLine(0,22)(0,110)

 \Vertex(0,22){2.5}
 \ArrowLine(-22,0)(0,22)

 \Photon(88,110)(0,110){2}{15}
 \Photon(88,22)(110,0){2}{5}

 \Photon(0,22)(88,22){2}{8}
 \Photon(88,22)(88,110){2}{8}
 \Photon(110,132)(88,110){2}{3}

 \ArrowLine(104,132)(88,116)
 \Vertex(88,110){2.5}
 \ArrowLine(104,0)(88,16)
 \ArrowLine(-16,0)(0,16)
 \ArrowLine(-16,132)(0,116)

 \Text(-15,105)[lb]{$\bf \alpha$}
 \Text(95,105)[lb]{ $\bf \nu$}
 \Text(95,20)[lb]{  $\bf \mu$}
 \Text(-15,15)[lb]{ $\bf \beta$}

 \Text(-4,125)[lb]{ $\bf p_1$}
 \Text(-4,0)[lb]{   $\bf p_2$}
 \Text(75,125)[lb]{ $\bf p_3$}
 \Text(75,0 )[lb]{  $\bf p_4$}

 \Text(-32,132)[lt]{$\bf \bar d$}
 \Text(-32,18)[lt]{ $\bf u$}
 \Text(115,132)[lt]{$\bf W$}
 \Text(115,18)[lt]{ $\bf \gamma$}
 \Text(-15,65)[lb]{$\bf d$} 
 \Text(40,117)[lb]{$\bf \gamma$}
 \Text(92,65)[lb]{ $\bf W$}
 \Text(37,12)[lt]{ $\bf W$}

  \Text(40,90)[lb]{$d_0$}
  \Text(10,63)[lb]{$d_1$}
  \Text(65,63)[lb]{$d_{3c}$}
  \Text(40,30)[lb]{$d_2$}

  \Text(20,-10)[lb]{$T_3$, crossed}
\end{picture}
}}
\end{array}
\]
\caption{Process $u\bar{d} \to WA$. Box topologies $T_1$ and $T_3$.}
\label{topologyT1_3}
\end{figure}

 For the process $u {\bar d} \rightarrow WA$,
 the box diagrams for the topologies $T_1,T_3$ are shown in Fig.~\ref{topologyT1_3}. Like
 topologies $T_2,T_4$ they are of {\it direct} and {\it crossed} type.

For the defining function $J^{T_1}$, we have:
\bqa
i\pi^2 J^{T_1} =\mu^{4-n}\int d^n q~ \frac{2 (q+p_1+p_2) p_1}{d_0(\mw)d_1(\mup)d_2(0)d_{3d}(\mw)}.
\label{JT1def}
\eqa
Due to occurrence of the photon mass, $\mgm=0$, in the argument list of the expression for
$J_{\rm uni}$, Eq.(\ref{AnswerFirstSet}) fails, and a special limit 
\bqa
J^0_{\rm uni}(P^2_1,P^2_2;m_1,0,m_3,m_4)=\lim_{m_2\to 0}~J_{\rm uni}(P^2_1,P^2_2;m_1,m_2,m_3,m_4)
\label{J0unideclar}
\eqa
has to be used instead. This is given by Eq.~(\ref{AnswerSecondSet}).

For the direct function $J^{T_1}$ the list of arguments of the universal function $J^0_{\rm uni}$
looks as follows:
\bqa
J^{T_1}=J^0_{\rm uni}(Q^2,T^2;\mw,0,\mw,\mup).
\label{JuniT1}
\eqa

For the cross channel topology $T_3$ the defining expression is:
\bqa
i\pi^2 J^{T_3}=\mu^{4-n}\int d^n q~ \frac{\ds -2 q  p_2}{d_0(0) d_1(\mdn) d_2(\mw)d_{3c}(\mw)},
\label{JT3def}
\eqa
and in terms of $J^0_{\rm uni}$ we have:
\bqa
J^{T_3}=J^0_{\rm uni}(Q^2,U^2;\mw,0,\mw,\mdn).
\label{JuniT3}
\eqa

{$\bullet$ $J^{T_1,T_3}$ as function of $D_0$ and $C_0$}

Performing the standard PV reduction, we express $J^{T_1,T_3}$
in terms of the corresponding $D_0$ and $C_0$ functions:
\bqa
J^{T_1} &=& \left(\Qs + \mw^2 \right) D_0( - \mdn^2, - \mup^2, - \mw^2, 0,\Qs,\Ts;\mw,\mup,0,\mw)
\nll &&
      - C_0( - \mup^2, - \mw^2,\Ts;\mup,0,\mw)  
      + C_0( - \mw^2, 0,\Qs;0,\mw,\mw),         
\label{relT1}
\eqa
and
\bqa
J^{T_3} &=& \left(\Qs+\mw^2\right)
       D_0(-\mdn^2,-\mup^2,-\mw^2,0,\Qs,\Us;\mw,\mup,0,\mw)
\nll &&
     - C_0( - \mw^2, - \mdn^2,\Us;\mw,0,\mdn)
     + C_0( 0, -\mw^2,\Qs;\mw,\mw,0).
\label{relT3}
\eqa
Again, relation (\ref{relT1}) holds exact in all masses involved, while relation (\ref{relT3}) holds
only up to quark mass power terms, $m^{2}_{u,d}/Q^{2}$, which we neglect.

$\bullet$ Subtracted functions $J^{T_{1,3}}_{\rm{sub}}$

Only one additional pinch has to be added to the relations~(\ref{relT1})--(\ref{relT3})
in order to cancel remaining mass singularities:
\bqa
J^{T_{1}}_{\rm{sub}} &=& \left(\Qs+\mw^2\right)
       D_0(-\mdn^2,-\mup^2,-\mw^2,0,\Qs,\Ts;\mw,\mup,0,\mw)
\nll &&
      -C_0(-\mup^2,-\mw^2,\Ts;\mup,0,\mw)
      +C_0(-\mw^2,0,\Qs;0,\mw,\mw)
\nll &&
      -\frac{\Qs}{\Ts+\mw^2} C_0(-\mup^2,-\mdn^2,\Qs;0,\mup,\mw),
\label{answT1sub}
\eqa
and
\bqa
J^{T_{3}}_{\rm{sub}} &=& \left( \Qs + \mw^2\right)
      D_0(-\mdn^2,-\mup^2,0,-\mw^2,\Qs,\Us;0,\mdn,\mw,\mw)
\nll &&
     -C_0(-\mw^2,-\mdn^2,\Us;\mw,0,\mdn)            
     +C_0(0,-\mw^2,\Qs;\mw,\mw,0)                
\nll &&
     -\frac{\Qs}{\Us+\mw^2} C_0(-\mup^2,-\mdn^2,\Qs;\mw,\mdn,0).
\label{answT3sub}
\eqa


\subsection{Pinches of topologies $T_1$ and $T_3$}
For topologies $T_1$ and $T_3$ we show only those three pinch diagrams which enter in the expessions
(\ref{answT1sub})--(\ref{answT3sub}) and give the explicit expressions for only one infrared-finite and
mass-singular pinch $C_{0,2}^{T_{1}}$. The function $C_{0,1}^{T_{1}}$ is
infrared and mass regular.

\vspace*{.5cm}

$\bullet$ {Topology $T_1$ pinches}

\vspace*{1cm}

\begin{figure}[!h]
\[
\begin{array}{ccc}
  \vcenter{\hbox{
\begin{picture}(88,88)(0,0)
\ArrowLine(22,110)(0,132)
\ArrowLine(0,66)(22,110)
\ArrowLine(-22,44)(0,66)
\Photon(22,110)(44,132){2}{5}
\Photon(0,66)(44,66){2}{8}
\Photon(44,66)(66,44){2}{3}
\Photon(44,66)(22,110){2}{4}
\Text(-10,125)[lb]{$\bf \bar d$}
\Text( 17,50)[lb]{ $\bf \gamma$}
\Text(48,125)[lb]{ $\bf \gamma$}
\Text(36,84)[lb]{  $\bf W$}
\Text(-7,84)[lb]{  $\bf u$}
\Text(-28,55)[lb]{ $\bf u$}
\Text(58,55)[lb]{  $\bf W$}
\Text(12,30)[lb]{$C_{0,\rm{IRD}}^{T_1}$}
\end{picture}
}}~~&
  \vcenter{\hbox{
\begin{picture}(88,88)(0,0)
\ArrowLine(-22,66)(0,88)
\ArrowLine(0,88)(-22,110)
\Photon(44,66)(44,110){2}{5}
\Photon(44,66)(66,44){2}{3}
\Photon(44,110)(0,88){2}{5}
\Photon(44,110)(66,132){2}{5}
\Photon(0,88)(44,66){2}{9}
\Text(45,85)[lb]{  $\bf W$}
\Text(-10,60)[lb]{ $\bf u$}
\Text(-10,110)[lb]{$\bf \bar d$}
\Text(10,60)[lb]{  $\bf \gamma$}
\Text(10,105)[lb]{ $\bf W$}
\Text(58,55)[lb]{  $\bf W$}
\Text(60,115)[lb]{ $\bf \gamma$}
\Text(12,30)[lb]{$C_{0,1}^{T_1}$}
\end{picture}
}}~~&
  \vcenter{\hbox{
\begin{picture}(88,88)(0,0)
\ArrowLine(0,66)(0,110)
\Photon(44,88)(0,110){2}{5}
\Photon(44,88)(66,66){2}{3}
\Photon(0,66)(44,88){2}{9}
\Photon(44,88)(66,110){2}{5}
\ArrowLine(-22,44)(0,66)
\ArrowLine(0,110)(-22,132)

\Text(-20,60)[lb]{ $\bf u$}
\Text(-20,110)[lb]{$\bf \bar d$}
\Text(22,60)[lb]{  $\bf \gamma$}
\Text(17,110)[lb]{ $\bf W$}
\Text(58,55)[lb]{  $\bf W$}
\Text(-20,82)[lb]{ $\bf u$}
\Text(60,115)[lb]{ $\bf \gamma$}
\Text(22,30)[lb]{$ C_{0,2}^{T_1}$}
\end{picture}
}}
\end{array}
\]

\vspace*{-1.25cm}

\caption{Diagrams of pinches for box topology ~$T_1$.}
\label{topologyTIpinch}
\end{figure}
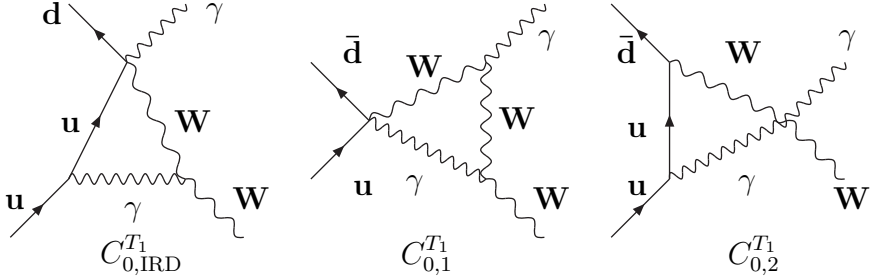
\vspace*{-.75cm}
\bqa
C_{0,\rm{IRD}}^{T_1}&=& C_0(-\mup^2,-\mw^2,\Ts;\mup,0,\mw),
\nll
C_{0,1}^{T_1}   &=& C_0(-\mw^2,0,\Qs;0,\mw,\mw),
\nll
C_{0,2}^{T_1}   &=& C_0(-\mup^2,-\mdn^2,\Qs;0,\mup,\mw)
\nll          &=& \frac{1}{\Qs}\Biggl[\ln\left(\frac{-\Qs}{\mw^2}\right)
                       \ln\left(\frac{\Qs+\mw^2-\ieps}{\mw^2}\right)
        +\Litwo\left(\frac{\Qs+\mw^2-\ieps}{\mw^2}\right)-\zeta(2)\Biggr].
\label{pinchesT1}
\eqa

\vspace*{.5cm}

$\bullet$ {Topology $T_3$ pinches}
\vspace*{1.5cm}

\begin{figure}[!h]
\[
\begin{array}{cccc}
  \vcenter{\hbox{
\begin{picture}(88,88)(0,0)
\ArrowLine(0,110)(-22,132)
\Photon(44,110)(0,110){2}{9}
\Photon(22,66)(44,110){2}{5}
\ArrowLine(22,66)(0,110)
\Photon(44,110)(66,132){2}{3}
\ArrowLine(0,44)(22,66)
\Photon(22,66)(44,44){2}{5}
\Text(-10,125)[lb]{$\bf \bar d$}
\Text(18,117)[lb]{ $\bf \gamma$}
\Text(40,125)[lb]{ $\bf W$}
\Text(38,84)[lb]{  $\bf W$}
\Text(-5,84)[lb]{  $\bf      d$}
\Text(-3,55)[lb]{  $\bf u$}
\Text(40,55)[lb]{  $\bf \gamma$}
\Text(5,30)[lb]{$C_{0,\rm{IRD}}^{T_3}$}
\end{picture}
}}~~~
&
  \vcenter{\hbox{
\begin{picture}(88,88)(0,0)

\ArrowLine(-22,66)(0,88)
\ArrowLine(0,88)(-22,110)

\Photon(0,88)(44,66){2}{5}%
\Photon(44,66)(44,110){2}{5}%
\Photon(44,110)(66,132){2}{3}
\Photon(44,66)(66,44){2}{5}
\Photon(44,110)(0,88){2}{7}%
\Text(-13,60)[lb]{ $\bf u$}
\Text(-12,110)[lb]{$\bf \bar d$}
\Text(12,62)[lb]{  $\bf W$}
\Text(17,107)[lb]{ $\bf \gamma$}
\Text(60,48)[lb]{  $\bf \gamma$}
\Text(48,82)[lb]{  $\bf W$}
\Text(60,115)[lb]{ $\bf W$}
\Text(12,30)[lb]{$C_{0,1}^{T_3}$}
\end{picture}
}}~~~
&
  \vcenter{\hbox{
\begin{picture}(88,88)(0,0)
\ArrowLine(0,66)(0,110)
\Photon(0,66)(44,88){2}{5}
\Photon(44,88)(0,110){2}{9}
\Photon(44,88)(66,66){2}{5}
\Photon(44,88)(66,110){2}{3}
\ArrowLine(-22,44)(0,66)
\ArrowLine(0,110)(-22,132)
\Text(-20,60)[lb]{ $\bf u$}
\Text(-20,110)[lb]{$\bf \bar d$}
\Text(17,60)[lb]{  $\bf W$}
\Text(22,110)[lb]{ $\bf \gamma$}
\Text(60,55)[lb]{  $\bf \gamma$}
\Text(-20,82)[lb]{ $\bf      d$}
\Text(60,115)[lb]{ $\bf W$}
\Text(17,30)[lb]{$C_{0,2}^{T_3}$}
\end{picture}
}}
\end{array}
\]

\vspace*{-1.25cm}

\caption{Diagrams of pinches for box topology ~$T_3$.}
\label{topologyTIIIpinch}
\end{figure}
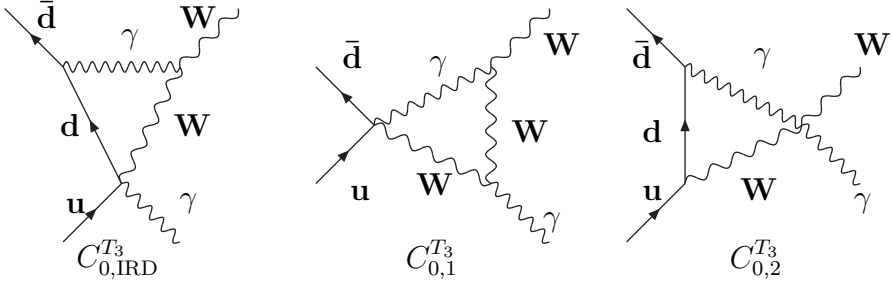
\vspace*{-.75cm}
\bqa
C_{0,\rm{IRD}}^{T_3} &=&  C_0(-\mw^2,-\mdn^2,\Us;\mw,0,\mdn),
\nll
C_{0,1}^{T_3}  &=&  C_0(0,-\mw^2,\Qs;\mw,\mw,0),
\nll
C_{0,2}^{T_3}  &=&  C_0(-\mup^2,-\mdn^2,\Qs;\mw,\mdn,0).
\label{pinchesT3}
\eqa
The explicit expression for pinch $C_{2}^{T_{3}}$ is obtained from $C_{2}^{T_{1}}$ by the
replacements $\Ts\to \Us$ and $\mdn\leftrightarrow\mup$.

\subsection{The final manipulations with functions $J^{T_1,T_3}_{\rm{sub}}$\label{finalmanT13}}
 Here the same argumentation applies as at the beginning of Section~\ref{finalmanT24},
but now for functions $J^{T_1,T_3}_{\rm{sub}}$. Here we have only one mass-singular pinch
$C_{0,2}^{T_1,T_3}$,
(\ref{pinchesT1}),~(\ref{pinchesT3}), and the equation, analogous to (\ref{answT2expl}), reads:
\bqa
J^{T_{1,3}}_{\rm{sub}} &=& J^{T_{1,3}} - \frac{\Qs}{\Ts+\mw^2} C_{0,2}^{T_{1,3}}.
\label{answT13expl}
\eqa
The functions $J^{T_1,T_3}$ and $C_{0,2}^{T_1,T_3}$ are substituted, the limits in the masses $\mdn\to 0$ and
$\mup\to 0$ are calculated, and we arrive at a rather compact answer:
\bqa
J^{T_1,T_3}_{\rm {sub}}(\Qs,\Ps,\mw^2) &=&
         \frac{1}{\Qs + \mw^2}
           \Biggl[-\Litwo\left(-\frac{\Qs}{\mw^2-\ieps}\right)+\zeta(2) \Biggr]
\nll &&
        +\frac{1}{\Ps+\mw^2}
   \Biggl\{
 \ln\left(-\frac{\mw^2-\ieps}{\Ps}\right) \ln\left(\frac{\Ps+\mw^2}{\mw^2}\right)
\nll &&
 +2\ln\left(\frac{\Qs+\mw^2-\ieps}{\mw^2}\right) \ln\left(\frac{\Ps+\mw^2}{\mw^2}\right)
\nll &&
  -\ln\left(\frac{\Qs+\mw^2-\ieps}{\mw^2}\right) \ln\left(\frac{-\Qs}{\mw^2}\right)
\nll &&
           -\Litwo\left(\frac{\Qs+\mw^2-\ieps}{\mw^2}\right)
           -\Litwo\left(\frac{\Ps+\mw^2}{\mw^2-\ieps}\right)+\zeta(2)
   \Biggr\}.
\label{answT1T3sim}
\eqa
Here again the list of physical arguments is restored and
$\Ps=\Ts$ for topology $T_1$ and $\Ps=\Us$ for topology $T_3$.
Similarly to~Eq.~(\ref{answT2T4sim}) we limit ourselves
to the $s$-channel kinematics -- where $\Qs < 0$ and $\Ps > 0$ -- by presenting the final
expression~Eq.~(\ref{answT1T3sim}).

On the other hand, one may invert Eqs.(\ref{answT1sub}) and (\ref{answT3sub}) in order to get rid of
the infrared divergent $D_0$, followed by redistribution of $C_0$'s as described at the end of
Section~\ref{finalmanT24}.

\newpage

\section{Topologies $T_6,T_{6'}$\label{Topology66}}

\subsection{Definition of functions $J^{T_6,T_{6'}}$}

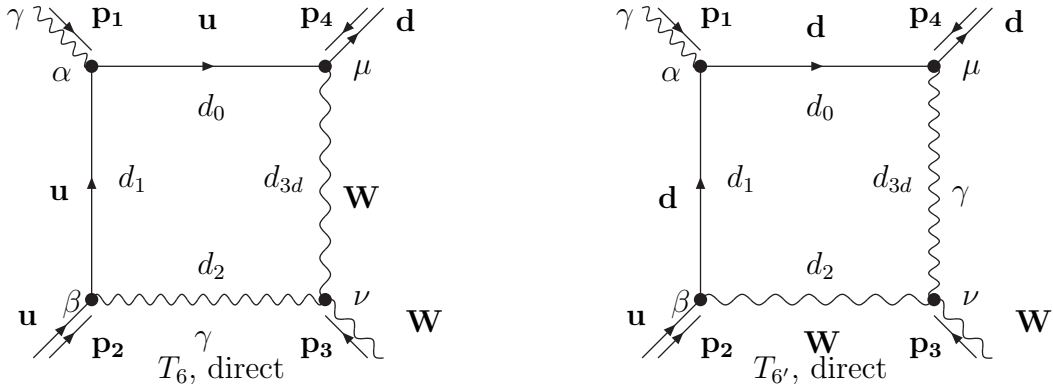
\begin{figure}[!h]
\[
\begin{array}{ccc}
  \vcenter{\hbox{
\begin{picture}(132,132)(0,0)
 \Text(-32,132)[lt]{$\bf \gamma$}
 \Text(-32,18)[lt]{ $\bf      u$}
 \Text(115,132)[lt]{$\bf      d$} 
 \Text(115,18)[lt]{ $\bf     W$}
 \Text(-20,65)[lt]{ $\bf u$}
 \Text(35,10)[lt]{ $\bf  \gamma$}
 \Text(40,130)[lt]{$\bf u$}
\Text(95,65)[lt]{$\bf W$}
 \Photon(88,22)(88,110){2}{7}
 \ArrowLine(0,110)(88,110)
 \Photon(0,22)(88,22){2}{11}
 \Vertex(88,22){2.5}
 \Photon(88,22)(110,0){2}{3}
 \ArrowLine(88,110)(110,132) 
 \Vertex(0,110){2.5}
 \Photon(0,110)(-22,132){2}{5}
 \ArrowLine(0,22)(0,110)
 \Vertex(0,22){2.5}
 \ArrowLine(-22,0)(0,22)
 \ArrowLine(-16,132)(0,116)
 \ArrowLine(-16,0)(0,16)
 \ArrowLine(104,132)(88,116)
 \Vertex(88,110){2.5}
 \ArrowLine(104,0)(88,16)
 \Text(-15,105)[lb]{$\bf \alpha$}
 \Text(95,105)[lb]{ $\bf \mu$}
 \Text(95,20)[lb]{  $\bf \nu$}
 \Text(-15,15)[lb]{ $\bf \beta $}
 \Text(-4,125)[lb]{ $\bf p_1$}
 \Text(-4,0)[lb]{   $\bf p_2$}
 \Text(75,125)[lb]{ $\bf p_4$}
 \Text(75,0 )[lb]{  $\bf p_3$}

  \Text(40,90)[lb]{$d_0$}
  \Text(10,63)[lb]{$d_1$}
  \Text(65,63)[lb]{$d_{3d}$}
  \Text(40,30)[lb]{$d_2$}

 \Text(25,-10)[lb]{$T_6$, direct}
\end{picture}
}}
& \qquad \qquad \qquad &
  \vcenter{\hbox{
\begin{picture}(132,132)(0,0)
 \Text(-32,132)[lt]{$\bf \gamma$}
 \Text(-32,18)[lt]{ $\bf      u$}
 \Text(115,132)[lt]{$\bf      d$} 
 \Text(115,18)[lt]{ $\bf      W$}
 \Text(-20,65)[lt]{ $\bf      d$} 
 \Text(35,10)[lt]{ $\bf       W$}
 \Text(40,130)[lt]{$\bf       d$} 
 \Text(95,65)[lt]{$\bf  \gamma$}
 \Photon(88,22)(88,110){2}{11}
 \ArrowLine(0,110)(88,110)
 \Photon(0,22)(88,22){2}{7}
 \Vertex(88,22){2.5}
 \Photon(88,22)(110,0){2}{3}
 \ArrowLine(88,110)(110,132)
 \Vertex(0,110){2.5}
 \Photon(0,110)(-22,132){2}{5}
 \ArrowLine(0,22)(0,110)
 \Vertex(0,22){2.5}
 \ArrowLine(-22,0)(0,22)
 \ArrowLine(-16,132)(0,116)
 \ArrowLine(-16,0)(0,16)
 \ArrowLine(104,132)(88,116)
 \Vertex(88,110){2.5}
 \ArrowLine(104,0)(88,16)
 \Text(-15,105)[lb]{$\bf \alpha$}
 \Text(95,105)[lb]{ $\bf \mu$}
 \Text(95,20)[lb]{  $\bf \nu$}
 \Text(-15,15)[lb]{ $\bf \beta $}
 \Text(-4,125)[lb]{ $\bf p_1$}
 \Text(-4,0)[lb]{   $\bf p_2$}
 \Text(75,125)[lb]{ $\bf p_4$}
 \Text(75,0 )[lb]{  $\bf p_3$}

  \Text(40,90)[lb]{$d_0$}
  \Text(10,63)[lb]{$d_1$}
  \Text(65,63)[lb]{$d_{3d}$}
  \Text(40,30)[lb]{$d_2$}

 \Text(20,-10)[lb]{$T_{6'}$, direct}
\end{picture}}}
\end{array}
\]
\caption{Process $u\bar{d} \to WA$. Box topologies $T_6$ and $T_{6'}$.}
\label{topologyT6_6pr}
\end{figure}

 For the process $u {\bar d} \rightarrow WA$,
 the box diagrams for the topologies $T_6,T_{6'}$ are shown in Fig.~\ref{topologyT6_6pr}.
 They both are of {\it direct} type (differing only by interchange of virtual $\gamma\leftrightarrow W$)
and hence relations of the type~(\ref{relT2}) and~(\ref{relT1}) will hold exactly in all masses.


For the defining function $J^{T_6}$, we have:
\bqa
i\pi^2 J^{T_6} =\mu^{4-n}\int d^nq \frac{2(q+p_1+p_2) p_1}{d_0(\mup)d_1(\mup)d_2(0)d_{3d}(\mw)}.
\label{JT6def}
\eqa
The function $J^{T_6}$ is expressed via the universal function $J^0_{\rm uni}$ by
\bqa
J^{T_6}=J_{\rm uni}(\Us,\Ts;\mw,\mdn,\mup,\mup).
\label{JuniT6}
\eqa
For $J^{T_{6'}}$, the pair of equations~(\ref{JT6def})--(\ref{JuniT6}) becomes:
\bqa
i\pi^2 J^{T_{6'}}=
\mu^{4-n} \int d^nq \frac{\ds 2(q+p_1+p_2+p_3)(-p_1)}{\ds d_0(\mdn)d_1(\mdn)d_2(\mw)d_{3d}(0)},
\label{JT6prdef}
\eqa
and in terms of the universal function
\bqa
J^{T_{6'}}=J_{\rm uni}(\Ts,\Us;\mw,\mup,\mdn,\mdn).
\label{JuniT6pr}
\eqa

\newpage

{$\bullet$ $J^{T_6,T_{6'}}$ as function of $D_0$ and $C_0$}

After the standard PV reduction, the functions $J^{T_6,T_{6'}}$ are expressed
in terms of the corresponding $D_0$ and $C_0$ functions by the pair of equations
\bqa
 J^{T_6} &=& \left(\Us + \mup^2\right)
        D_0\left(0, - \mup^2, - \mw^2, - \mdn^2,\Us,\Ts;\mup,\mup,0,\mw \right)
\nll &&
      - C_0\left( - \mup^2, - \mw^2,\Ts;\mup,0,\mw \right)
      + C_0\left( - \mw^2, - \mdn^2,\Us;0,\mw,\mup \right),
\label{relT6}
\eqa
and
\bqa
J^{T_{6'}} &=& \left(\Ts + \mdn^2 \right)
        D_0\left(0, -\mup^2, -\mw^2, -\mdn^2,\Us,\Ts;\mdn,\mdn,\mw,0\right)
\nll &&
      - C_0\left( - \mw^2, - \mdn^2,\Us;\mw,0,\mdn\right)
      + C_0\left( - \mup^2, - \mw^2,\Ts;\mdn,\mw,0\right);
\label{relT6pr}
\eqa

$\bullet$ Subtracted functions $J^{T_{6,6'}}_{\rm{sub}}$

\bqa
J^{T_6}_{\rm{sub}} &=& \left(\Us + \mup^2\right)
        D_0\left(0, -\mup^2, -\mw^2, -\mdn^2,\Us,\Ts;\mup,\mup,0,\mw \right)
\nll &&
      - C_0(-\mup^2,-\mw^2,\Ts;\mup,0,\mw) + C_0(-\mw^2,-\mdn^2,\Us;0,\mw,\mup)
\nll &&
      - \frac{\Us}{\mw^2+\Ts} C_0(0,-\mup^2,\Us;\mup,\mup,0)
\nll &&
      - \frac{\Ts}{\mw^2+\Ts} C_0(0,-\mdn^2,\Ts;\mup,\mup,\mw),
\label{answT6sub}
\eqa
and
\bqa
J^{T_{6'}}_{\rm{sub}} &=& \left(\Ts + \mdn^2\right)
       D_0(0,-\mup^2,-\mw^2,-\mdn^2,\Us,\Ts;\mdn,\mdn,\mw,0)
\nll &&
      - C_0(-\mw^2,-\mdn^2,\Us;\mw,0,\mdn)
      + C_0(-\mup^2,-\mw^2,\Ts;\mdn,\mw,0)
\nll &&
      - \frac{\Ts}{\mw^2+\Us} C_0(0,-\mdn^2,\Ts;\mdn,\mdn,0)
\nll &&
      - \frac{\Us}{\mw^2+\Us} C_0(0,-\mup^2,\Us;\mdn,\mdn,\mw).
\label{answT6prsub}
\eqa
Here again all four pinches are present in the relations~(\ref{answT6sub})--(\ref{answT6prsub})
for the subtracted functions $J^{T_{6,6'}}_{\rm{sub}}$.

\newpage

\subsection{Pinches of topologies $T_6$ and $T_{6'}$}


$\bullet$ {Topology $T_6$ pinches}

\vspace*{1cm}

\begin{figure}[!h]
\[
\begin{array}{ccccc}
  \vcenter{\hbox{
\begin{picture}(88,88)(0,0)
\ArrowLine(-22,44)(0,66)
\Photon(0,66)(44,66){2}{11}
\ArrowLine(0,66)(22,110)
\ArrowLine(22,110)(44,132)
\Photon(22,110)(0,132){2}{7}
\Photon(44,66)(22,110){2}{5}
\Photon(44,66)(66,44){2}{3}
\Text(17,50)[lb]  {$\bf \gamma$}
\Text(48,125)[lb] {$\bf d$}
\Text(60,55)[lb]  {$\bf W$}
\Text(-3,88)[lb]  {$\bf u$}
\Text(-10,125)[lb]{$\bf \gamma$}
\Text(35,88)[lb]  {$\bf W$}
\Text(-20,55)[lb] {$\bf u$}
\Text(10,30)[lb]{$C_{0,\rm IRD}^{T_6}$}
\end{picture}}}  &
  \vcenter{\hbox{
\begin{picture}(88,88)(0,0)
\ArrowLine(-22,66)(0,88)     
\Photon(0,88)(-22,110){2}{5}
\ArrowLine(0,88)(44,110)     
\Photon(0,88)(44,66){2}{7}   
\Photon(44,66)(44,110){2}{5}
\ArrowLine(44,110)(66,132)
\Photon(44,66)(66,44){2}{3}
\Text(-10,60)[lb] {$\bf u$}
\Text(-10,110)[lb]{$\bf \gamma$}
\Text(15,60)[lb]  {$\bf \gamma$}
\Text(15,110)[lb] {$\bf u$}
\Text(60,55)[lb]  {$\bf W$}
\Text(50,88)[lb]  {$\bf W$}
\Text(60,115)[lb] {$\bf d$}
\Text(12,30)[lb]{$C_{0,1}^{T_6}$}
\end{picture}}}    &
  \vcenter{\hbox{
\begin{picture}(88,88)(0,0)
\ArrowLine(0,66)(0,110)
\Photon(0,66)(44,88){2}{7}
\ArrowLine(0,110)(44,88)
\Photon(44,88)(66,66){2}{3}
\ArrowLine(44,88)(66,110) 
\ArrowLine(-22,44)(0,66)  
\Photon(0,110)(-22,132){2}{5}
\Text(-20,60)[lb] {$\bf u$}
\Text(-20,110)[lb]{$\bf \gamma$}
\Text(22,60)[lb]  {$\bf \gamma$}
\Text(22,110)[lb] {$\bf u$}
\Text(60,55)[lb]  {$\bf W$}
\Text(-20,82)[lb] {$\bf u$}
\Text(60,115)[lb] {$\bf d$}
\Text(17,30)[lb]{$C_{0,2}^{T_6}$}
\end{picture}}}   &
  \vcenter{\hbox{
\begin{picture}(88,88)(0,0)
\Photon(0,110)(-22,132){2}{5}
\ArrowLine(0,110)(44,110)
\Photon(22,66)(44,110){2}{5}
\ArrowLine(22,66)(0,110)
\ArrowLine(44,110)(66,132)
\ArrowLine(0,44)(22,66)
\Photon(22,66)(44,44){2}{3}
\Text(-10,125)[lb]{$\gamma$}
\Text(22,117)[lb] {$\bf u$}
\Text(44,125)[lb] {$\bf d$}
\Text(40,82)[lb]  {$\bf W$}
\Text(-3,82)[lb]  {$\bf u$}
\Text(0,55)[lb]   {$\bf u$}
\Text(40,55)[lb]  {$\bf W$}
\Text(10,30)[lb]{$C_{0,3}^{T_6}$}
\end{picture}}}
\end{array}
\]

\vspace*{-1.5cm}

\caption{Diagrams of pinches for box topology ~$T_6$.}
\label{topologyTVIpinch}
\end{figure}

These pinches correspond to the following $C_0$ functions:
\bqa
C_{0,\rm{IRD}}^{T_6} &=& C_0(-\mup^2,-\mw^2,\Ts;\mup,0,\mw),
\nll
C_{0,1}^{T_6} &=& C_0(-\mw^2,-\mdn^2,\Us;0,\mw,\mup),
\nll
C_{0,2}^{T_6} &=& C_0(0,-\mup^2,\Us;\mup,\mup,0),
\nll
C_{0,3}^{T_6} &=& C_0(0,-\mdn^2,\Ts;\mup,\mup,\mw)
\nll        &=&\frac{1}{\Ts}\Biggl[\frac{1}{2}
 \ln^2\left(\frac{\Ts+\mw^2}{\mw^2}\right)
         +\ln\left(\frac{\Ts+\mw^2}{\mw^2}\right)
          \ln\left(\frac{\mw^2}{\mup^2}\right)
         -\Litwo\left(\frac{\Ts}{\Ts+\mw^2}\right)\Biggr].
\label{pinchesT6}
\eqa
The function $C_{0,1}^{T_6}$ is mass-singularity free,
the explicit expression for $C_{0,2}^{T_6}$ was given above in Eqs.(\ref{pinchesT2}) for $C_{0,2}^{T_2}$.
Here we give an explicit expression only for $C_{0,3}^{T_6}$, which was not presented so far.


\vspace*{.5cm}

$\bullet$ {Topology $T_{6'}$ pinches}

\vspace*{1cm}

\begin{figure}[!h]
\[
\begin{array}{ccccc}
  \vcenter{\hbox{
\begin{picture}(88,88)(0,0)
\Photon(44,66)(44,110){2}{7}
\Photon(0,88)(-22,110){2}{5}
\ArrowLine(-22,66)(0,88)
\ArrowLine(0,88)(44,110)
\Photon(0,88)(44,66){2}{5}
\ArrowLine(44,110)(66,132)
\Photon(44,66)(66,44){2}{3}
\Text(-10,60)[lb] {$\bf u$}
\Text(-10,110)[lb]{$\bf \gamma$}
\Text(15,60)[lb]  {$\bf W$}
\Text(16,107)[lb] {$\bf d$}
\Text(60,55)[lb]  {$\bf W$}
\Text(50,82)[lb]  {$\bf \gamma$}
\Text(60,115)[lb] {$\bf d$}
\Text(12,30)[lb]{$C_{0,\rm{IRD}}^{T_{6'}}$}
\end{picture}}}   &
  \vcenter{\hbox{
\begin{picture}(88,88)(0,0)
\ArrowLine(-22,44)(0,66)
\ArrowLine(0,66)(22,110)
\ArrowLine(22,110)(44,132)
\Photon(22,110)(0,132){2}{5}
\Photon(44,66)(22,110){2}{7}
\Photon(0,66)(44,66){2}{5}
\Photon(44,66)(66,44){2}{3}
\Text(17,50)[lb]  {$\bf W$}
\Text(48,125)[lb] {$\bf d$}
\Text(60,55)[lb]  {$\bf W$}
\Text(-3,88)[lb]  {$\bf d$}
\Text(-10,125)[lb]{$\bf \gamma$}
\Text(35,88)[lb]  {$\bf \gamma$}
\Text(-20,60)[lb] {$\bf u$}
\Text(15,30)[lb]{$C_{0,1}^{T_6}$}
\end{picture}}}   &
  \vcenter{\hbox{
\begin{picture}(88,88)(0,0)
\Photon(0,110)(-22,132){2}{5}
\ArrowLine(0,110)(44,110)
\Photon(22,66)(44,110){2}{7}
\ArrowLine(22,66)(0,110)
\ArrowLine(44,110)(66,132)
\ArrowLine(0,44)(22,66)
\Photon(22,66)(44,44){2}{3}
\Text(-10,125)[lb]{$\gamma$}
\Text(22,117)[lb] {$\bf d$}
\Text(44,125)[lb] {$\bf d$}
\Text(40,82)[lb]  {$\bf \gamma$}
\Text(-3,84)[lb]  {$\bf d$}
\Text(0,55)[lb]   {$\bf u$}
\Text(40,55)[lb]  {$\bf W$}
\Text(10,30)[lb]{$C_{0,2}^{T_6}$}
\end{picture}}}   &
  \vcenter{\hbox{
\begin{picture}(88,88)(0,0)
\ArrowLine(0,66)(0,110)
\Photon(0,66)(44,88){2}{5}
\ArrowLine(0,110)(44,88)
\Photon(44,88)(66,66){2}{3}
\ArrowLine(44,88)(66,110) 
\ArrowLine(-22,44)(0,66)  
\Photon(0,110)(-22,132){2}{5}
\Text(-20,60)[lb] {$\bf u$}
\Text(-20,110)[lb]{$\bf \gamma$}
\Text(22,60)[lb]  {$\bf W$}
\Text(22,110)[lb] {$\bf d$}
\Text(60,55)[lb]  {$\bf W$}
\Text(-20,82)[lb] {$\bf d$}
\Text(60,115)[lb] {$\bf d$}
\Text(22,30)[lb]{$C_{0,3}^{T_6}$}
\end{picture}}}
\end{array}
\]

\vspace*{-1.5cm}
\caption{Diagrams of pinches for box topology ~$T_{6'}$.}
\label{topologyTVIprpinch}
\end{figure}

These pinches correspond to the $C_0$ functions with arguments
interchanged as compared to~Eqs.(\ref{pinchesT6}):
\bqa
&& C_{0,\rm{IRD}}^{T_{6'}} = C_0(-\mw^2,-\mdn^2,\Us;\mw,0,\mdn),
\nll
&& C_{0,1}^{T_{6'}}   = C_0(-\mup^2,-\mw^2,\Ts;\mdn,\mw,0),
\nll
&& C_{0,2}^{T_{6'}}   = C_0(0,-\mdn^2,\Ts;\mdn,\mdn,0),
\nll
&& C_{0,3}^{T_{6'}}   = C_0(0,-\mup^2,\Us;\mdn,\mdn,\mw).
\label{pinchesT6pr}
\eqa
The explicit expression for pinch $C_{3}^{T_{6'}}$ is obtained from $C_{3}^{T_{6}}$
by the replacements $\Ts\to \Us$ and $\mdn\leftrightarrow\mup$.

\subsection{The final manipulations with functions $J^{T_6,T_{6'}}_{\rm{sub}}$\label{finalmanT56}}
The same argumentation as in the beginning of Section~\ref{finalmanT24} applies here.
The analogue of ``way one'' expressions~(\ref{answT2expl}) and~(\ref{answT13expl}) in this case reads:
\bqa
J^{T_6}_{\rm{sub}} &=& J^{T_6} - \frac{\Us}{\mw^2+\Ts} C_{0,2}^{T_6} - \frac{\Ts}{\mw^2+\Ts} C_{0,3}^{T_6}.
\label{answT6expl}
\eqa
After substitution of its ingredients and taking the limits $\mdn\to 0$ and $\mup\to 0$, we arrive
at a very short answer:
\bqa
J^{T_6}_{\rm {sub}}(\Ts,\Us,\mw) &=&
      \frac{1}{\mw^2+\Us}\left[\frac{1}{2}\ln^2\left(\frac{\mw^2}{\Us}\right)
                +3 \Litwo(1)\right]
\nll && +\frac{1}{\mw^2+\Ts}
 \Biggl[\frac{1}{2}\ln^2\left(\frac{\mw^2}{\Us}\right)
                      -\ln^2\left(\frac{\mw^2+\Ts}{\Us}\right)-3 \Litwo(1)\Biggr].
\label{answT6sim}
\eqa
Here the restored list of physical arguments corresponds to the topology $T_6$. The answer for topology
$T_{6'}$ obtains by interchange $\Ts\leftrightarrow \Us$.
The expression Eq.~(\ref{answT6sim}) for the $s$-channel kinematics where $\Ts > 0$ and $\Us > 0$
is real (has no imaginary part).

As usual, one may invert Eqs.(\ref{answT6sub}) and (\ref{answT6prsub}) in order to exclude the
infrared divergent $D_0$, followed by redistribution of $C_0$'s as described at the end of
Section~\ref{finalmanT24}.

\newpage

\section{Mass singularity free combinations of $D_0$ and $C_0$ functions\label{Combd0c0}}

The ``second way'', being applied to all six box topologies with a virtual photon line,
eventually leads to the cancellation of many but not all mass-singular $C_{0,i}^{T_{j}}$ functions.
The remaining mass-singular $C_0$ cancel after observation that certain linear combinations of $D_0$
(associated with boxes having a virtual $Z$ in place of a virtual photon)
and $C_0$ do not contain mass singularities. We verified numerically (with the aid of LoopTools) that
the four following combinations converge to a stable limit when $\mup\to 0$ and/or $\mdn\to 0$:
\bqa
C_{d_0,c_0}(\Qs,\Ts) &=&(\Qs\Ts+\Qs\mz^2+\mw^2\mz^2)
\nll &&
 \times D_0(-\mdn^2,-\mup^2,-\mw^2,0,\Qs,\Ts;\mdn,\mz,\mup,\mdn)
\nll &&
                               -\Ts C_0(0,-\mdn^2,\Ts;\mdn,\mdn,\mz)
\nll &&
                  -(\mw+\Qs)~C_0(0,-\mw^2,\Qs;\mdn,\mdn,\mup);
\nll
C_{d_0,c_0}(\Qs,\Us) &=&(\Qs\Us+\Qs\mz^2+\mw^2\mz^2)
\nll &&
  \times D_0( - \mdn^2, - \mup^2, 0, - \mw^2, \Qs,\Us;\mdn,\mz,\mup,\mup)
\nll &&
                  -\Us C_0(0,-\mup^2,\Us;\mup,\mup,\mz)
\nll &&
                  -(\Qs+\mw)~C_0(0,-\mw^2,\Qs;\mup,\mup,\mdn);
\nll
C_{d_0,c_0}(\Ts,\Us) &=&(\Ts\Us+\mw^2\Us+\mz^2\Ts)
\nll &&
 \times D_0(0,-\mup^2,-\mw^2,-\mdn^2,\Us,\Ts;\mup,\mup,\mz,\mw)
\nll &&
                  -\Us C_0(0,-\mup^2,\Us;\mup,\mup,\mz)
\nll &&
                  -\Ts C_0(0,-\mdn^2,\Ts;\mup,\mup,\mw);
\nll
C_{d_0,c_0}(\Us,\Ts) &=&(\Ts\Us+\mw^2\Us+\mz^2\Ts)
\nll &&
 \times D_0(0,-\mup^2,-\mw^2,-\mdn^2,\Us,\Ts;\mdn,\mdn,\mw,\mz)
\nll &&
                 -\Ts C_0(0,-\mdn^2,\Ts;\mdn,\mdn,\mz)
\nll &&
                 -\Us C_0(0,-\mup^2,\Us;\mdn,\mdn,\mw).
\label{Combinations}
\eqa
Note the nontrivial kinematical coefficients in front of the $D_0$ functions.

Use equations (\ref{Combinations}) to exclude four $D_0$ functions in favour of
of the mass-singular function $C_0$ and of $C_{d_0,c_0}$ which are
free of mass singularities. One can verify that all 12 mass-singular $C_0$
functions cancel in the complete expression for the NLO PW part of the cross-section of the process
under consideration, $ud\to WA$. A wonderful fact is that the mass-singular $C_0$ cancel as a
whole, i.e. without substituting their explicit expressions.

\section{Numerical Results}
\label{NumRes}

In this Section we compare the real and imaginary 
parts of the function $J^{T_{i}}_{\rm{sub}}$, presented in Sections \ref{finalmanT24}, \ref{finalmanT13}
and \ref{finalmanT56}, with the corresponding ones, computed using the LoopTools package~\cite{Hahn:1998yk}.

In the Tables below, SANC results are presented in the first rows, and the
corresponding LoopTools numbers in the second rows.

The numbers are given for two values of $s$ (in GeV$^2$) and for three values of $\cos\vartheta$.

\begin{table}[!h]
{\small
\begin{tabular}{||c|c|c|c|c||}
\hline
\hline
 $\cos\vartheta$  & \multicolumn{2}{|c|}{${s} = 10^4$} & \multicolumn{2}{|c|}{${s} = 10^6$}    \\
\hline
       & Re                  & Im                   & Re                   & Im             \\
\hline
-0.999 & 6.53638473617E-08   & -1.13107515511E-07   &  3.89750326994E-11   & -3.12226209474E-09 \\
       & 6.53638447452E-08   & -1.13107515739E-07   &  3.89750340292E-11   & -3.12226209489E-09 \\ 
\hline
 0     & 9.73334338213E-05   & -1.24690402239E-04   &  5.48800329682E-07   & -4.31508792787E-06 \\
       & 9.73334338175E-05   & -1.24690402239E-04   &  5.48800329683E-07   & -4.31508792787E-06 \\
\hline
 0.999 & 8.28985241530E-04   & -2.80233766861E-04   &  5.40166075955E-05   & -3.12695141375E-05 \\ 
       & 8.28985239959E-04   & -2.80233766861E-04   &  5.40166075954E-05   & -3.12695141375E-05 \\
\hline
\hline
\end{tabular}
}
\caption{
Comparison of real and imaginary parts of function $J^{T_{2}}_{\rm{sub}}$ between SANC and LoopTools results
calculated for different values of $s$ and $\cos\vartheta$. The mass of $M_{\sss W} =$ 80 GeV. 
For the topology $T_4$ the rows $\pm 0.999$ have to be interchanged.
}
\label{Tableone}
\end{table}
As is seen from Table~\ref{Tableone}, there is agreement from 7 to 12 digits for real and 
imaginary parts.

\begin{table}[!h]
{\small
\begin{tabular}{||c|c|c||}
\hline
\hline
 $\cos\vartheta$  &   $s = 10^4$         &     $s = 10^6$       \\
\hline
               &  Re                  &   Re                \\
\hline
 -0.999        & -2.86644118212E-04  &  -2.43416713242E-03  \\ 
               & -2.86644118211E-04  &  -2.43416713242E-03  \\
\hline
  0            & -8.41693567906E-05  &   3.76366826830E-05  \\
               & -8.41693567907E-05  &   3.76366826830E-05  \\
\hline
  0.999        &  1.40586637158E-03  &   1.12954414152E-03  \\ 
               &  1.40586637160E-03  &   1.12954414152E-03  \\
\hline
\end{tabular}
}
\caption{
Comparison of the
real function $J^{T_{6}}_{\rm{sub}}$ between SANC and LoopTools results
calculated with different values $s$ and $\cos\vartheta$; $M_{\sss W} =$ 80 GeV.
For the topology $T_{6'}$ the rows $\pm 0.999$ have to be interchanged.
}
\label{Tabletwo}
\end{table}
As is seen from Table~\ref{Tabletwo}, we have  again agreement 
within 10-12 digits for 
the functions $J^{T_{6,6'}}_{\rm{sub}}$, which are real for topologies $T_6$ and $T_{6'}$.


 \begin{table}[!h]
{\small
\begin{tabular}{||c|c|c|c|c||}
\hline
\hline
 $\cos\vartheta$       & \multicolumn{2}{|c|}{$s = 10^4$} &   \multicolumn{2}{|c|}{$s = 10^6$}\\
\hline
         & Re                 & Im                & Re                 & Im                \\
\hline
 -0.999  &  1.85671149365E-04 & 2.50678493204E-04 &  1.42521459789E-05 & 9.82990632508E-08 \\
         &  1.85671149365E-04 & 2.50678493205E-04 &  1.42521459789E-05 & 9.82990632508E-08 \\
\hline
  0      &  2.36366656601E-04 & 3.69642886168E-04 &  3.20290753820E-05 &-6.95783750976E-06 \\
         &  2.36366656601E-04 & 3.69642886168E-04 &  3.20290753820E-05 &-6.95783750976E-06 \\
\hline
  0.999  &  3.27769575491E-04 & 5.99387751646E-04 & -1.50514580606E-03 & 2.22442249799E-03 \\
         &  3.27769575491E-04 & 5.99387751646E-04 & -1.50514580606E-03 & 2.22442249799E-03 \\
\hline
\hline
\end{tabular}
}
\caption{
Comparison of real and imaginary parts of function $J^{T_{1}}_{\rm{sub}}$ between SANC and LoopTools results
calculated for different values $s$ and $\cos\vartheta$. The mass of $M_{\sss W} =$ 80.4 GeV.
For the topology $T_3$ the rows $\pm 0.999$ have to be interchanged.
}
\label{Tablethree}
\end{table}

As is seen from the Table~\ref{Tablethree}, there is agreement from 11 to 12 digits for real and
imaginary parts for the topologies $T_1$ and $T_3$.

\vspace{0.5cm}

The numerical comparison with the LoopTools library presented 
in this paper can be verified with help of the SANC software packages. 

We have made a package related to the functions $J_{T_i}$, called JAWAudWA.F.
This is  available to download from the web page of project SANC~\cite{homepagesSANC}.

\clearpage

\section{Conclusions}
In this paper we continue the study of the
infrared and mass singularities emerging from 4-point function box diagrams 
with an internal photon line connecting two external
lines on the mass shell, on the example of the process $ud \to WA$.

Here we extend our earlier investigations of the calculation of diagrams of
such a class:
see~\cite{Bardin:1999ak} and ~\cite{Bardin:2009zz},
where the general approach to this problem was originally proposed.

The approach consists of three steps.
In the first step we introduce a new class of auxiliary functions $J$, relevant to the 
Passarino--Veltman reduction~\cite{Passarino:1978jh} of scalar and vector integrals. By construction,
$J$ functions are free of infrared singularities and are made sufficiently simple for subsequent
integration over the three Feynman parameters $z$, $x$, $y$, leading to a compact explicit result
in terms of dilogarithm functions. The function $J$, in turn, may be subjected to the standard 
Passarino-Veltman reduction giving linear combinations of the standard $D_0$ and $C_0$ functions,
which may be used to exclude complicated infrared divergent $D_0$ function in favour of $J$ function
and simplest 3-point infrared divergent $C_0$ function.  
 
In general, the explicit form of $J$ function is not universal,
depending on the concrete topology of the infrared divergent $D_0$ function of a process.

There are six different topologies of the infrared divergent box diagrams which are met 
in the analytic calculations of functions $J$ for process $ud \to WA$. 
For this case we found a way to introduce a universal function by means of a special trick to
simplify the analytic calculations, choosing two 4-vectors and Feynman parameterization 
in the defining expression for functions $J$, Eq.~(\ref{JuniGen}), which
ensures linear dependence of the integrand of $J$ on one of the integration variables, $x$,
Eq.~(\ref{JAWint3}).

In this way,
we received the expression for $J$ in terms of the universal auxiliary function
$J_{\rm uni}(P^2_1,P^2_2;m_1,m_2,m_3,m_4)$.
This allows us to obtain explicit expression for various topologies by a simple rotation 
of its dummy arguments. This is new compared to our previous papers on $J$ functions.

The second step is typical for the SANC treatment of $J$ functions:
for each $J_{\rm uni}$ we build the corresponding subtracted $J_{\rm{sub}}$ functions free
of mass singularities, which are shifted to some other set of mass singular $C_0$ functions.
A part of the latter $C_0$ functions cancels at this step.

The third step consists of combining the four remaining mass singular $D_0$ functions with all
remaining mass singular $C_0$ functions. These combinations $C_{d_0,c_0}$, Eq.~(\ref{Combinations}),
are again free of mass singularities, and all 12 mass singular $C_0$ functions of the problem cancel. 

This approach leads to compact analytical results, allows one
to perform stable and fast numerical calculations
and avoid large numerical cancellations between separate terms.



\bigskip\noindent
{\bf Acknowledgements.}
This work is partly supported by Russian Foundation for Basic Research
grant $N^{o} 12-02-91526-{\rm CERN\_a}$.


\bibliographystyle{utphys_spires}
\addcontentsline{toc}{section}{\refname}
\bibliography{JAW_udWA}

\end{document}